%% file: sfb676_B1.tex
\documentclass{sfb676}

\makeatletter
\g@addto@macro\bfseries{\boldmath}
\makeatother

\input{sfb676_macros_B1.tex}

\begin{document}
\title{Optimising the ILC Setup: Physics Programme, Running Scenarios and Design Choices}

\author{{\slshape Jenny List$^1$, Gudrid Moortgat-Pick$^{1,2}$, J\"urgen Reuter$^1$}\\[1ex]
$^1$DESY, Hamburg, Germany\\
$^2$II. Institut f\"ur Theoretische Physik, Universit\"at Hamburg, Germany}

\contribID{B1}

\desyproc{PUBDB-2018-00782}
\acronym{SFB 676 -- Particles, Strings and the Early Universe} 
\doi  

\maketitle

\begin{abstract}
A high-energy $e^+e^-$ Linear Collider has been considered since a long time as an important complement to the LHC. Unprecedented precision measurements as well as the exploration of so far untouched phase space for direct production of new particles will provide unique information 
to advance the limits of our understanding of our universe. 
Within this project, the physics prospects of such a collider as well as their interplay with design of the accelerator and the detectors have been investigated in a quantitative way. This kind of study required a close collaboration between theory and experiment, always taking into account results of the LHC and other relevant experiments. 
In this article we will summarize some of the most important developments and results, covering all core areas of the physics progamme of future $e^+e^-$ colliders.
\end{abstract}

%
%
%

\section{Introduction} 


A high energy electron positron collider enables unprecedented precision studies of the Higgs boson, the top quark, the heavy gauge bosons and possibly yet unknown particles, and would thus be an ideal complement to the LHC.  From its beginning, the SFB676-B1 project has been dedicated to the quantitative assessment of the physics case of such a future $e^+e^-$ 
collider~\cite{Moortgat-Picka:2015yla}, considering existing LHC results as well as still to be expected measurements from the HL-LHC. As a concrete example the most advanced of such projects, namely the International Linear Collider (ILC)~\cite{Behnke:2013xla}, has been used in order to study the impact of accelerator parameters and detector performance on the physics prospects~\cite{Berggren:2010wy}. In many cases, the assessment was based on realistic technology assumptions verified e.g.\ with prototypes in testbeam. 

\begin{figure}[tb]
\centering
  \begin{subfigure}{.485\linewidth}
    \includegraphics[width=\textwidth]{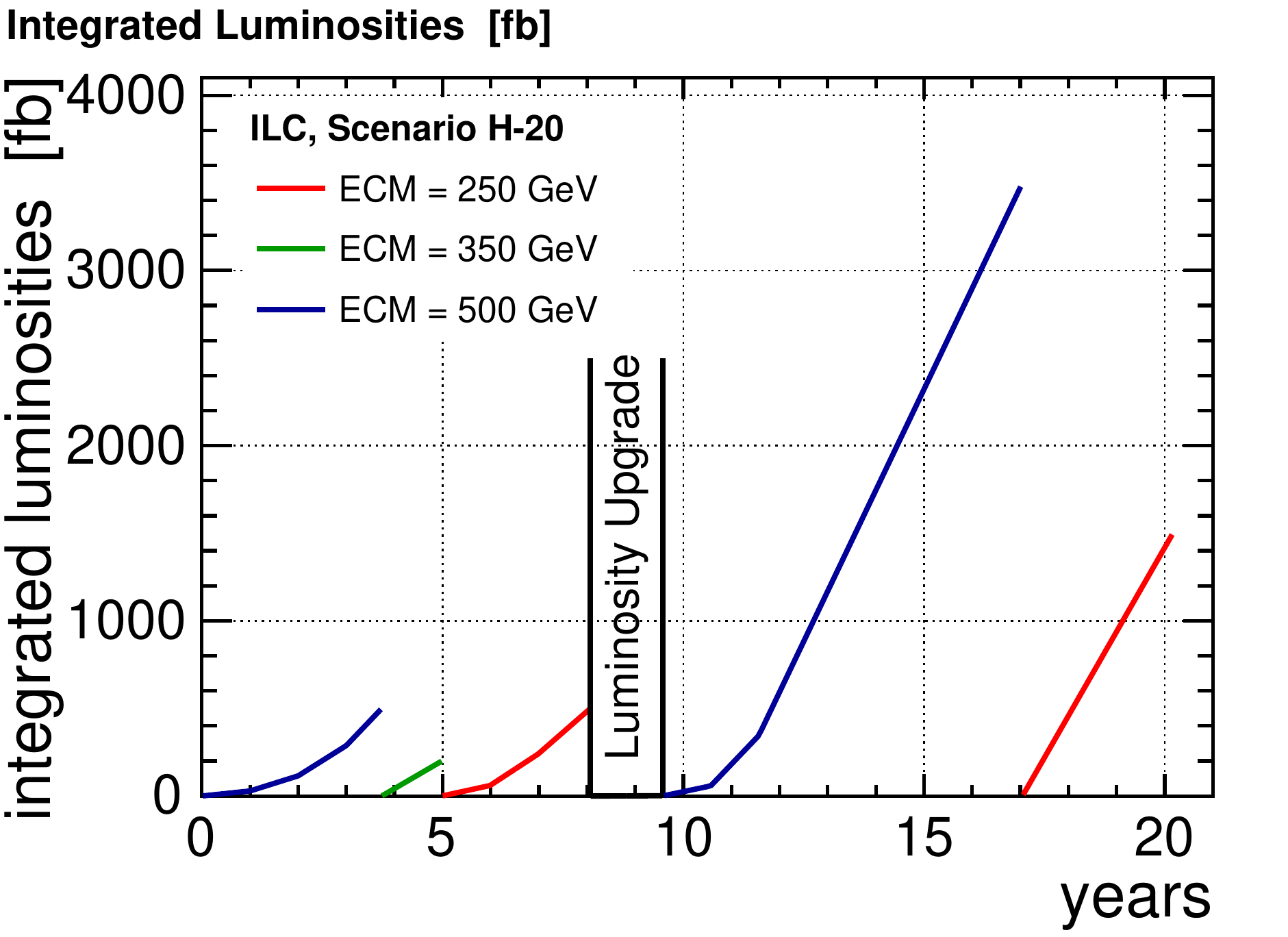}
    \caption{}
    \label{fig:lumiH20}
  \end{subfigure}
	\hfill
  \begin{subfigure}{.485\linewidth}
    \includegraphics[width=\textwidth]{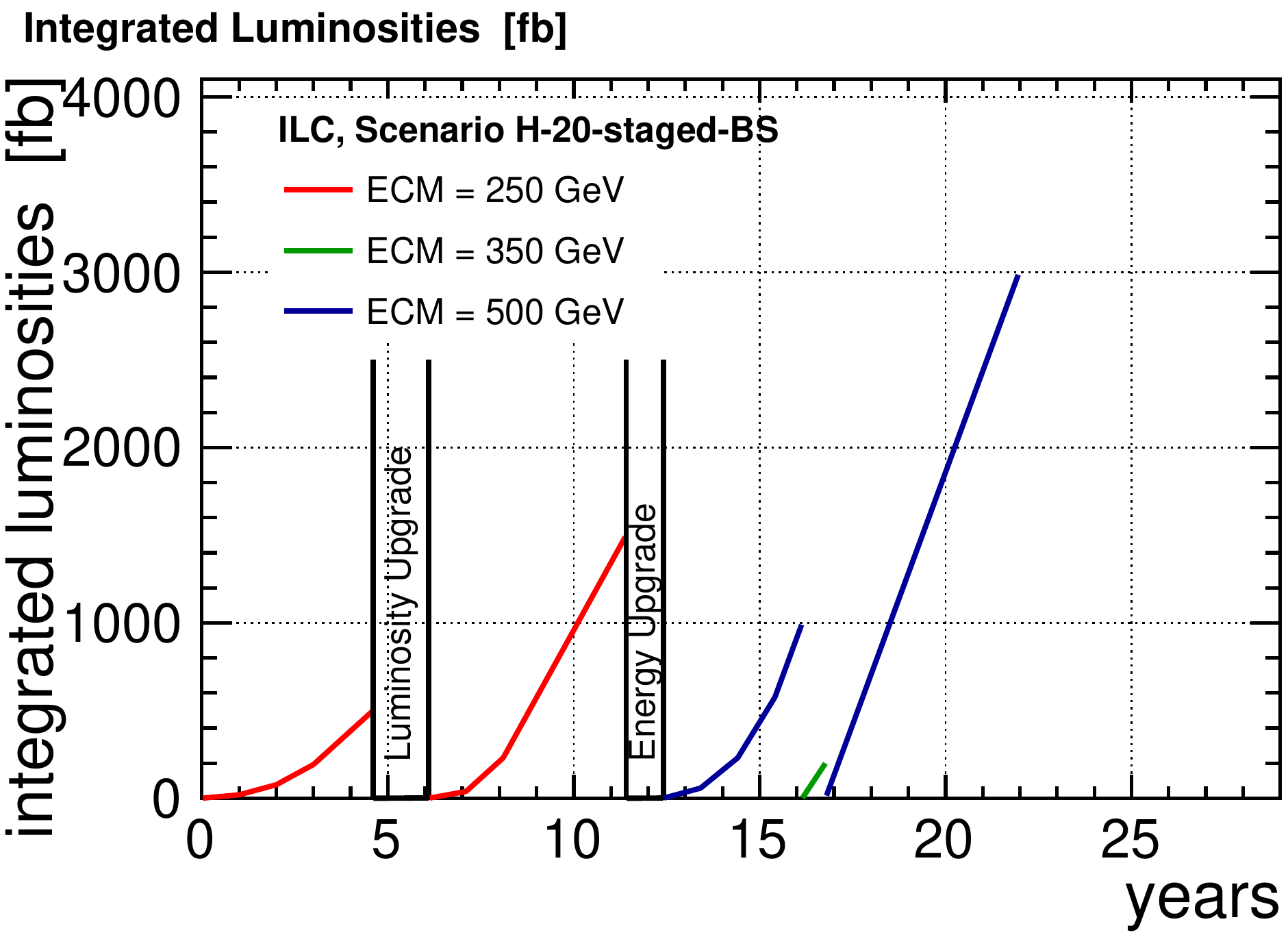}
    \caption{}
    \label{fig:lumiH20stagedBS}
  \end{subfigure}
  \caption{Integrated luminosity vs time in two construction scenarios for the ILC (a) immediate construction of a 500\,GeV machine (b) staged construction starting with a 250\,GeV collider. Figures taken from Refs.~\cite{Barklow:2015tja} and~\cite{Fujii:2017vwa}, respectively.}
  \label{fig:runscen}
\end{figure}

The actual discovery of the Higgs boson at the LHC in 2012, near the mid-term of this project, gave a tremendous boost to the planning of such a collider. With the knowledge of the Higgs mass being close to 125\,GeV, the energy thresholds of important processes like single or double Higgsstrahlung could be pin-pointed with certainty for the first time. As a result, the Linear Collider Collaboration (LCC) defined concrete operating scenarios for the ILC~\cite{Barklow:2015tja}, with  strong involvement of this SFB project. The total amount of integrated luminosity collected at each energy stage were defined such as to optimize the expected precision on the couplings of the Higgs boson for a total operation period of about 20 years. The official default running scenario is shown in Fig.~\ref{fig:lumiH20}, which serves as a reference for all ILC physics studies since. An optimal early physics performance is achieved when starting operation at a center-of-mass energy of 500\,GeV, runs at $250$\,GeV are very important for ultimate precision on the coupling of the Higgs to the $Z$ boson. The initial construction costs, however, are much lower when starting at the lowest possible energy, because only a part of the accelerating modules is needed at the beginning. Therefore, also a staged version of the default running scenario has been developed~\cite{Fujii:2017vwa}, again with leading contributions from this project. The staged scenario, which after the full program delivers the same integrated luminosities as Fig.~\ref{fig:lumiH20}, is 
shown in Fig.~\ref{fig:lumiH20stagedBS}.

Besides the center-of-mass energy and the luminosity, the polarisation of the electron and positrons beams is an important top-level parameter of electron-positron colliders. The impact of beam polarisation on the physics program of future $e^+e^-$ colliders has been an important topic throughout the lifetime of this project, and has been summarized  in~\cite{MoortgatPick:2010zza, MoortgatPick:2010zz}. Also the precise monitoring of the luminosity-weighted average beam polarisation from collision data received several imporant contributions from this project~\cite{LIST:2014yga, marchesini11:_tgcs_and_pol_at_ILC_and_leakage_in_calo, Karl:2017xra}. In particular a global fit framework has been developed which can for the first time combine total and differential cross-section measurements from many different physics processes as well as the polarimeter measurements~\cite{Karl:2017xra}. This framework has been used to study the ultimately achievable precisions for each dataset in the H20 running scenario, as well as to study the impact of systematic uncertainties and their correlations. Most recently, the role of positron polarisation has been reviewed in view of the staging proposal for the ILC~\cite{Fujii:2018mli}, again with leading contributions from this project.

All the studies of the physics potential of a future $e^+e^-$ collider which will be summarized in the following sections rest on advanced software tools from MC generators over detector simulations to reconstruction and data analysis. Nearly all studies for future linear colliders world-wide rely on the MC Generator {\sc Whizard}~\cite{Kilian:2007gr}. In close collaboration with the project B11 of this SFB, strong contributions have been made the automation of NLO QCD corrections in {\sc Whizard}, 
but also to the implementation of Linear Collider specific features and requirements~\cite{Kilian:2014nya}. The latter profited strongly from the direct collaboration between {\sc Whizard} developers and users 
within the SFB. 

Algorithmic developments to event reconstruction at Linear Colliders comprise e.g.\ a novel technique to account for initial state radiation and Beamstrahlung in kinematic fits~\cite{Beckmann:2010ib, Beckmann:2010zz}. Prior to this development the benefit of kinematic fits was considered to be much smaller at Linear Colliders than e.g.\ at LEP due to the much higher radiation losses.
With the new technique, $e^+e^- \to W^+W^- / ZZ \to 4$\,jet-events with significant photon radiation could be fitted equally well as events without radiation, achieving comparable resolutions e.g.\ on di-jet masses. This technique was applied successfully since in several ILC and CLIC analyses, e.g.\ in a PhD thesis written in context of this project on the prospects for measuring the triple-Higgs coupling at the ILC~\cite{Duerig:2016dvi}. 

In the following sections, we will highlight results obtained in this project in the some of main areas of the physics program of future $e^+e^-$ colliders: Precision studies of the Higgs boson and searches for its potential siblings, precision studies of 
the $W$ and $Z$ bosons, as well as seaches for the direct production of new particles.

\section{Higgs physics in the SM and beyond} 
With the discovery of the Higgs boson the questions of the inner workings of electroweak symmetry breaking and of the stabilisation of the Higgs mass became unavoidable. Precision measurements of the Higgs boson's properties provide unique key information for solving these puzzles. The SFB-B1 project contributed in many aspects to the quantitative understanding of the possible measurements and to their interpretation. These will be summarized in this section together with  the experimentally closely related search for siblings of the Higgs boson with reduced couplings to the $Z$ boson.

\subsection{Higgs couplings to fermions and bosons}

One of the unique opportunities at $e^+e^-$ colliders is the determination of the total decay width of the Higgs boson. This can be achieved either by measuring the total Higgsstrahlungs cross section and the subsequent decay of the Higgs into a pair of $Z$ bosons -- or by studying Higgs production in $WW$ fusion with subsequent decay of the Higgs into a pair of $W$ bosons. Since the former is limited by the small branching ratio of $H \to ZZ$, the latter has been studied in~\cite{Durig:2014lfa}, considering especially dependency of the achievable precision on the center-of-mass energy. It was shown that without making any assumptions on the relations between the couplings of the Higgs to $W$ and $Z$ bosons, e.g.\ by custodial symmetry, data-taking at center-of-mass energies of at least $350$\,GeV, better $500$\,GeV is mandatory.

The prospects and challenges of measuring the Higgs branching ratios into pairs of $b$- or $c$-quarks or of gluons has been studied at a center-of-mass energy of $350$\,GeV~\cite{Mueller:2016exq}. At this energy, the Higgsstrahlung production mode with subsequent decay of the $Z$ into two neutrinos and $WW$ fusion production mode contribute about equally. Within this project it was shown for the first time that 
all three hadronic decay modes as well as the two production channels can be disentangled simultaneously by using not only the flavour tag information but also the invariant mass of the missing four-momentum as discriminating variables.  

\begin{figure}[tb]
\centering
  \begin{subfigure}{.485\linewidth}
    \includegraphics[width=\textwidth]{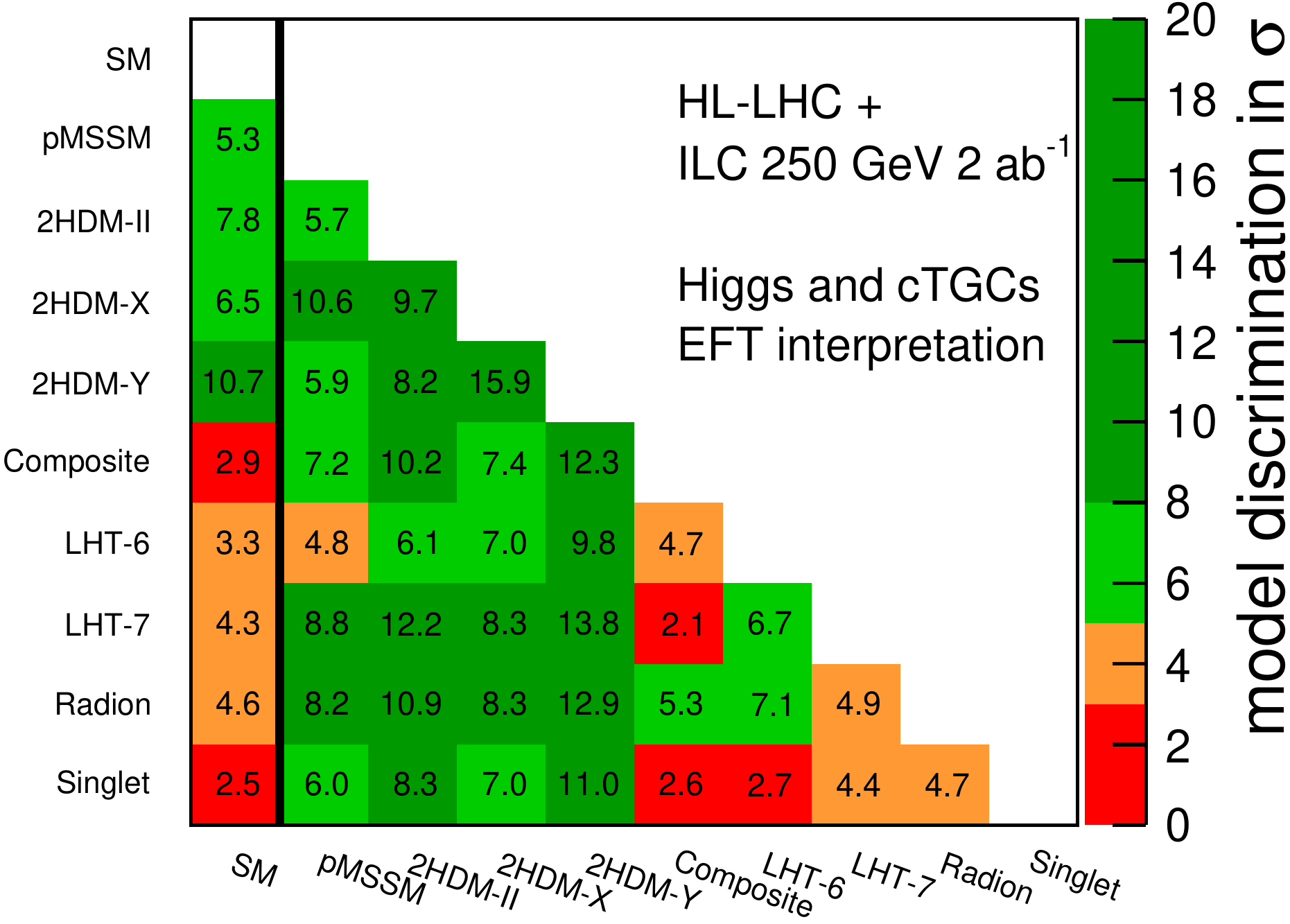}
    \caption{}
    \label{fig:triangle250}
  \end{subfigure}
	\hfill
  \begin{subfigure}{.485\linewidth}
    \includegraphics[width=\textwidth]{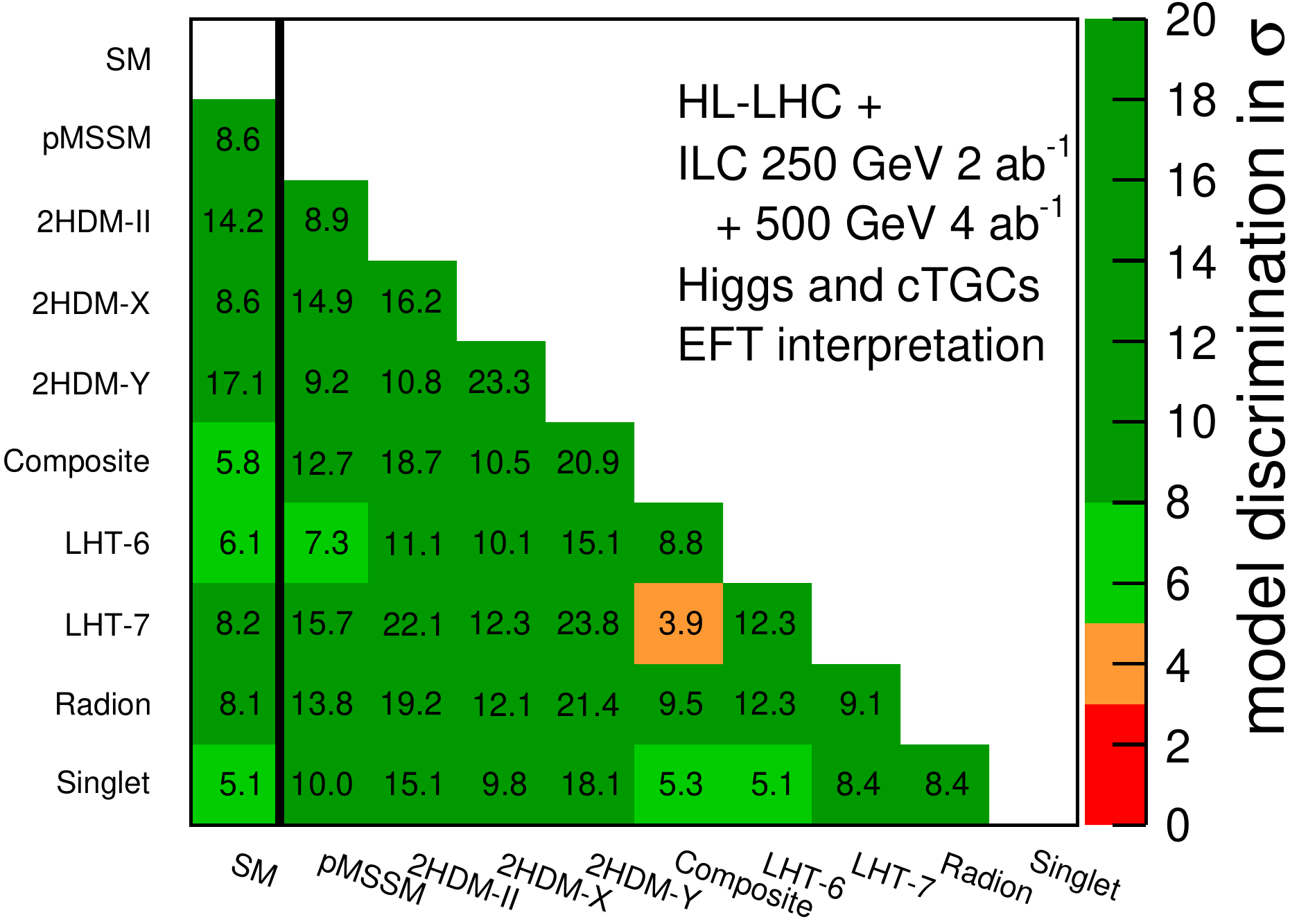}
    \caption{}
    \label{fig:triangle500}
  \end{subfigure}
\caption{Discrimination power between the SM and various examplatory BSM benchmarks from ILC Higgs and electroweak precision measurements. All benchmarks have been chosen to be unobservable at the (HL-)LHC. (a) For the 250\,GeV stage of the ILC (b) Adding the full 500\,GeV program. Figures taken from Ref.~\cite{Barklow:2017suo}.}
\label{fig:bsmchi2}
\end{figure}

 
These results as well as many more which have been obtained within the world-wide ILC community have been used as input to the definition of default running scenarios for the ILC mentioned in the Introduction. While these were still based on a $\kappa$-framework type of interpretation, a new approach was implemented in 2017 within an effective operator framework considering all dim-6 operators consistent with $SU(2)\times U(1)$ symmetry~\cite{Barklow:2017suo}. In such a framework, the symmetry assumptions as well as the inclusion of triple gauge coupling constraints allow to constrain the total decay width of the Higgs boson much better than in the $\kappa$-framework already at $\sqrt{s}=250$\,GeV. A special contribution from the SFB-B1 project is shown in Fig.~\ref{fig:bsmchi2}, which illustrates the power of Higgs (and electroweak) precision measurements for discriminating various new physics benchmark scenarios from the SM and from each other, based on the EFT-interpretation of projected ILC measurements. All benchmark points have been chosen such that they will not be observable at the HL-LHC, their definition can be found in~\cite{Barklow:2017suo}. Fig.~\ref{fig:triangle250} displays the discrimination power in terms of the number of standard deviations for the 250\,GeV ILC, while Fig.~\ref{fig:triangle500} corresponds to the full H20 program, including the 500\,GeV data.
 
Another important measurement is the direct determination of the Higgs self-coupling $\lambda$ from double-Higgs production.  High-energy Linear Colliders offer two very complementary opportunities for this measurement: At energies of about 1\,TeV and higher, pairs of Higgs bosons can be produced in vector boson fusion, where the cross section decreases with larger values of $\lambda$, like in the analogous process at hadron colliders. At energies around 500\,GeV, double Higgs-strahlung is accessible, whose cross section grows with increasing $\lambda$.  A PhD thesis written in the context of the SFB-B1 project showed for the first time the feasibility of this measurement at the ILC with $\sqrt{s}=500$\,GeV in full detector simulation and including pile-up from soft photon-photon collisions~\cite{Duerig:2016dvi}.


\subsection{Additional light Higgs bosons}


Beyond the precision study of the 125\,GeV Higgs boson, the $e^+e^-$ colliders also offer unique possibilities to search for additional light scalars $S^0$ with reduced couplings to the $Z$ boson, be it additional Higgs bosons or
PNGBs which occur frequently in various extensions of the SM. The special handle here is to use the same recoil method which also ensures the model-independent determination of the total Higgsstrahlungs cross section. Thereby, the four-momentum of the $S^0$ is reconstructed solely from the decay products of the $Z$ boson, e.g.\ in $Z \to \mu^+ \mu^-$ and $Z\to e^+e^-$, and the known initial state. Within the SFB-B1 project, the sensitivity of the 250\,GeV ILC to such additional scalars has been studied in an interdisciplinary endeavour in two approaches:
a) by extrapolating the analogous LEP searches~\cite{Drechsel:2018mgd}, as well as  
b) in full simulation of the ILD detector concept~\cite{Wang:2018fcw}. 

\begin{figure}[tb]
\centering
  \begin{subfigure}{.44\linewidth}
    \includegraphics[width=\textwidth]{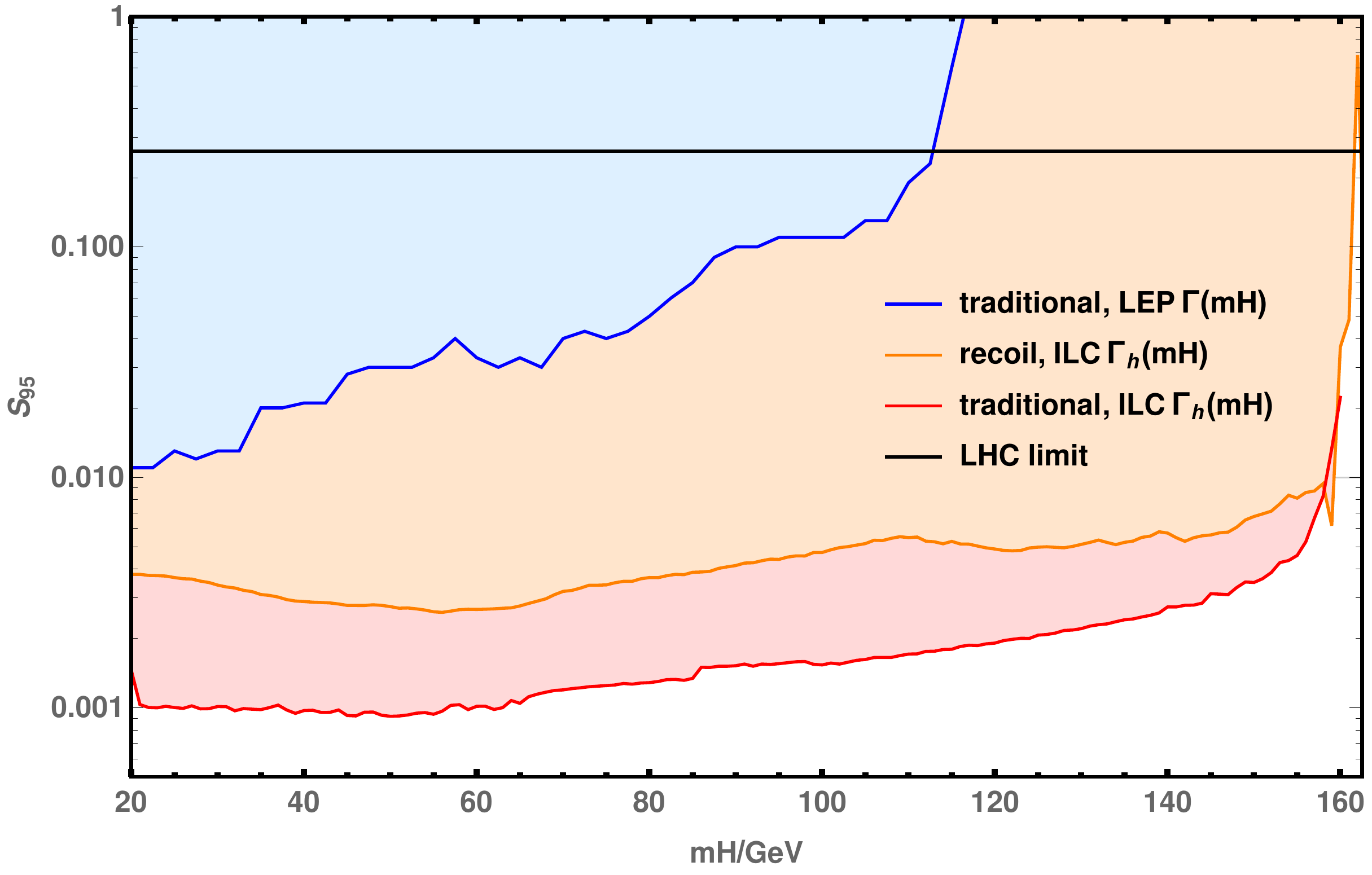}
    \caption{}
  \label{fig:lighthiggsDrechsel}
  \end{subfigure}
	\hfill
  \begin{subfigure}{.55\linewidth}
    \includegraphics[width=\textwidth]{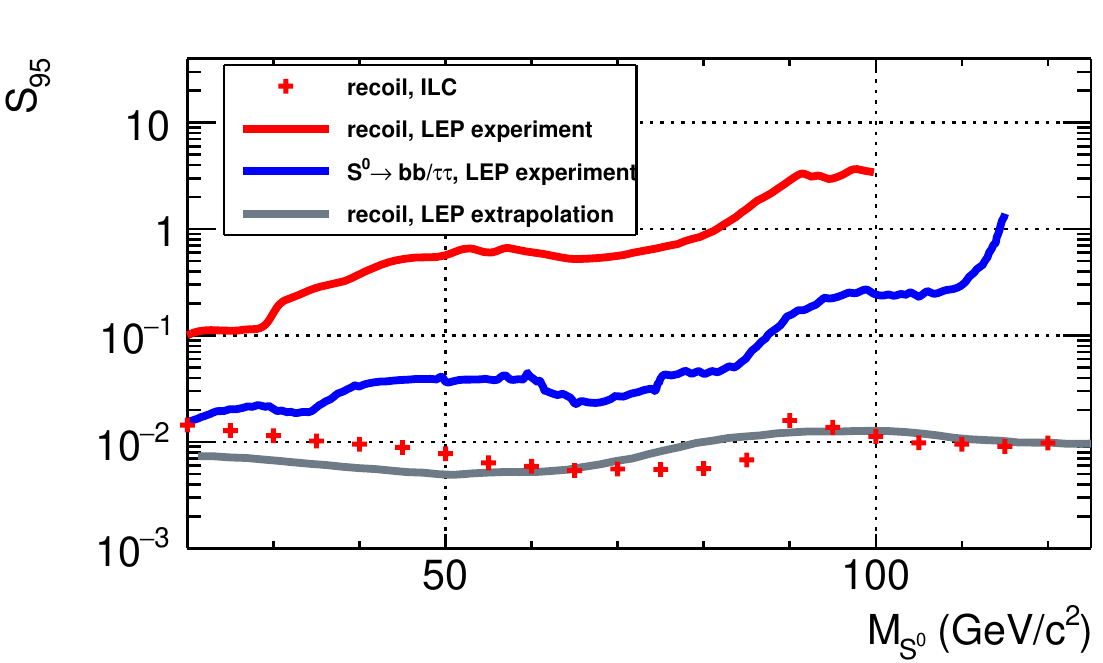}
    \caption{}
  \label{fig:lighthiggsWang}
  \end{subfigure}
  \caption{Expected sensitivity at 95\% C.L. as a function of the mass of the $S^0$ in terms of the signal cross section normalized to the SM Higgsstrahlung cross section for the same mass at the ILC with $\sqrt{s}=250$\,GeV:
(a)  Decay-mode independent recoil analysis in the process $e^+e^- \to HZ$, $Z\to \mu^+\mu^-$, $e^+e^-$ as well as
the decay-mode dependent analysis in the process $e^+e^- \to HZ\to b \bar{b} \mu^+\mu^-$ with ${\cal L}=500$~fb$^{-1}$
 and  $(P_{e^-},P_{e^+})=(-80\%,+30)$; method checked against LEP data analysis and extrapolated to ILC.
(b)  Decay-mode independent recoil analysis in the process $e^+e^- \to HZ$, $Z\to \mu^+\mu^-$. The red crosses show the reach of the ILC in full detector simulation for an integrated luminosity of 2000\,fb$^{-1}$ and 
$(P_{e^-},P_{e^+})=(\mp 80\%,\pm 30\%)$ [$(\mp 80\%,\mp 30\%)$] in the luminosity ratio $40\%$ [10\%], respectively. The red line shows the directly comparable exclusion limit from LEP, the blue line a decay mode-dependent search from LEP. The grey line corresponds to the yellow line from panel (a).
Figures taken from Ref.~\cite{Drechsel:2018mgd} and~\cite{Wang:2018fcw}, respectively. }
  \label{fig:lighthiggs}
\end{figure}

At LEP such light Higgses have been studied in two ways: i) via analyzing both decaying particles explicitly, $Z\to \mu^+ \mu^-$ and $H\to b \bar{b}$  (called 'LEP traditional'), and ii) via applying a Higgs-decay independent method, the recoil method, and analyzing only the decays $Z\to \mu^+ \mu^-, e^+ e^-$. 
After a reproducing the LEP sensitivities as a verification of the method, both LEP searches were extrapolated to ILC energy, luminosity and polarization, i.e.\ for $\sqrt{s}=250$~GeV, ${\cal L}=500$~fb$^{-1}$ and the polarization configuration $(P_{e^-},P_{e^+})=(-80\%,+30\%)$~\cite{Drechsel:2018mgd}.
The resulting expected limit on $S$ at the 95\% C.L.\ can be seen in Fig.~\ref{fig:lighthiggsDrechsel}.  
Following the LEP convention, the sensitivity is expressed as a ratio of the cross section for $Z S^0$ production over the (by now hypothetical) SM $Z H$ cross section for the same mass, called $S$. 
This ratio is proportional to the squares of the involved couplings $S \sim \left| g^2_{S^0 ZZ}/g^2_{HZZ} \right|$.

\begin{figure}[tb]
\centering
    \includegraphics[width=.6\textwidth]{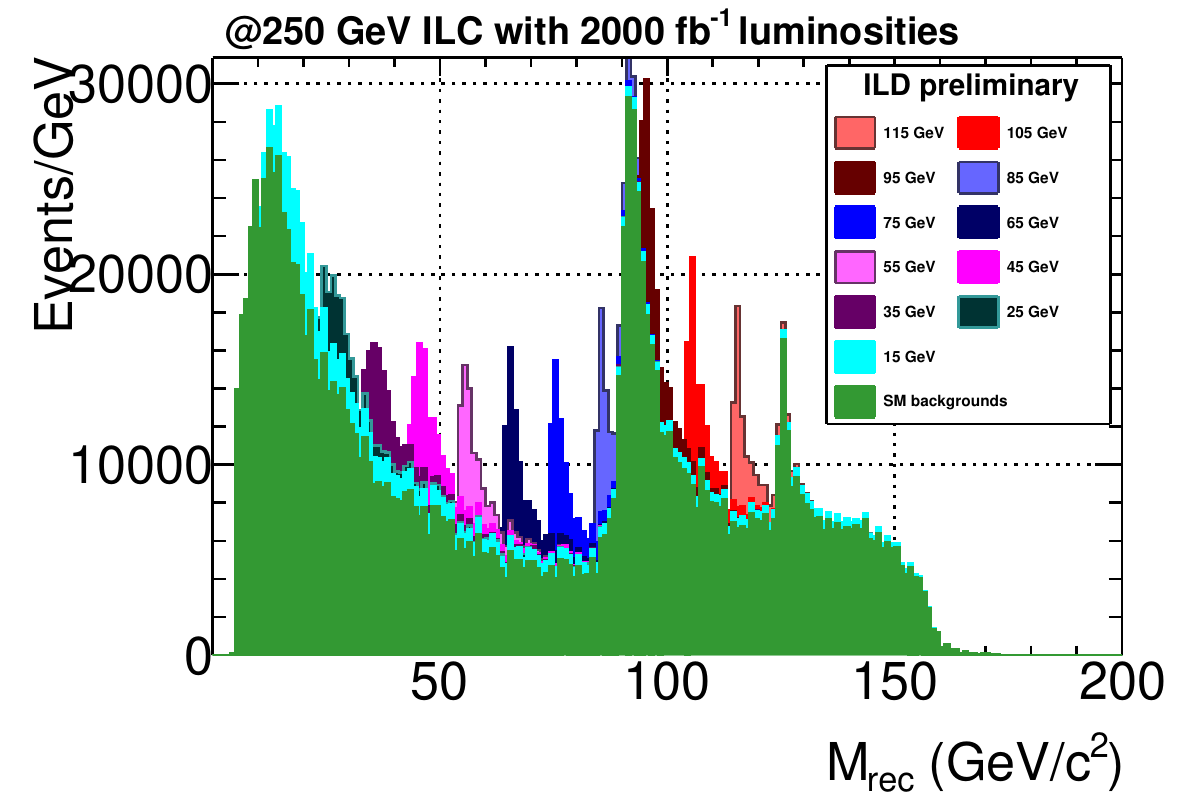}
  \caption{Search for additional light scalars: SM backgrounds superimposed with $S^0$ signals for various masses in full simulation of the ILD concept~\cite{Wang:2018fcw}.}
     \label{fig:mrecoil_lighthiggs}
\end{figure}

The channel $e^+e^- \to S^0Z$, $Z\to \mu^+\mu^-$ was studied in
in full simulation of the ILD detector concept for ${\cal L}=2000$~fb$^{-1}$ splitted up into the four polarization configurations $(P_{e^-},P_{e^+})=(\mp 80\%,\pm 30\%)$ [$ (\mp 80\%,\mp 30\%)$] with 40\% [10\%] of the total luminosity, respectively~\cite{Wang:2018fcw}.
Figure~\ref{fig:mrecoil_lighthiggs} shows the resulting recoil mass spectrum from all SM backgrounds (note that this includes the 125\,GeV Higgs) and $S^0$ signals of different masses with an arbitrary normalisation. These spectra were then be used to project
sensitivities on $S$ at 95\% C.L.\  as a function of the $S^0$ mass,  shown in Fig.~\ref{fig:lighthiggsWang}.


While the two analyses are broadly consistent, there are some differences in the results: for very light Higgs masses they originate from different approaches in treating the width of the $S^0$, while at higher masses the different considered final states of the $Z$-boson play a role.

These studies show that the ILC covers significant additional parameter space down to about a 1\% of the SM $ZH$ cross section in a decay-mode independent way over a broad range of light Higgs masses, even at a center-of-mass energy of only 250\,GeV. 

\subsection{Off-shell effects and the Higgs width}

The exploitation of off-shell contributions in Higgs processes can play an important role to determine properties of the Higgs particles.
In \cite{Liebler:2015aka} the off-shell contributions in $H\to V V^*$ with $V={Z,W}$ have been studied. Both dominant
production processes $e^+e^-\rightarrow ZH\rightarrow ZVV^{(*)}$ and
$e^+e^-\rightarrow \nu\bar\nu H\rightarrow \nu\bar\nu VV^{(*)}$ 
are taken into account. The relative size of the off-shell contributions is strongly
dependent on the centre-of-mass energy. These
contributions can have an important impact on the
determination of cross sections and branching ratios. Furthermore, the combination of on- and off-shell contributions 
can be exploited to test higher-dimensional operators,
unitarity and light and heavy Higgs interferences in extended Higgs.
sectors.

The fact that the mass of the observed Higgs boson of about $125$\,GeV
is far below the threshold for on-shell $W^+W^-$ and $ZZ$ production has the
consequence that the decay $H\rightarrow VV^{*}$ of an on-shell Higgs
boson suffers from a significant phase-space suppression. This implies on
the one hand that the partial width $H\rightarrow VV^{*}$, where $H$ is
on-shell, depends very sensitively on the precise numerical value of the 
Higgs-boson mass. On the other hand, contributions of an off-shell Higgs
with  decays into two on-shell $VV$  are relatively large. The relative importance of
contributions of an off-shell Higgs boson increases with increasing
cms energy, cf. Table\ref{table:higgs-gmp}. For $\sqrt{s}>500$\,GeV those off-shell contributions 
to the total Higgs induced cross section are of
$\mathcal{O}(10\%)$. 

The extraction of Higgs couplings to gauge bosons from 
branching ratios of $H\rightarrow VV^{*}$ require a very precise measurement
of the Higgs-boson mass (preferably better than $100$\,MeV).
At low cms energies~$\sqrt{s}$,
i.e.\ close to the production threshold, the effects of 
off-shell contributions are insignificant for the extraction of
Higgs couplings. For an accurate determination of Higgs couplings
at higher $\sqrt{s}$, however, the off-shell contributions
have to be incorporated. 

A particular focus of our analysis has been on the determination of the
total width of the Higgs boson at a linear collider. We have investigated
two aspects in this context. On the one hand, we have analysed to what
extent the standard method at a linear collider, which is based on the 
$Z$ recoil method providing an absolute measurements of Higgs 
branching ratios in combination with an appropriate determination of a
partial width, is affected by off-shell contributions. We have found that
at low cms energies the effect of the off-shell
contributions in $H\rightarrow VV^{(*)}$ is at the sub-permil level.
At higher energies, however, the off-shell effects are larger and need to
be properly taken into account and/or reduced by appropriate cuts. 
However, the method based on the comparison of on-shell and off-shell
contributions has several draw-backs. Besides relying heavily on theoretical 
assumptions, this method requires very high statistics and is limited by 
the negative interference term. We therefore conclude that the standard
method at a linear collider based on the 
$Z$ recoil method is far superior for determining the Higgs
width, both because of its model-independence and the
much higher achievable precision. We have also discussed the corresponding
method at the LHC and we have pointed out that the destructive
interference contribution between the Higgs-induced contributions and the
background will make it difficult to reach the sensitivity to
the SM value of the width even for high statistics.

As an example of the relevance of off-shell effects in the context of an
extended Higgs sector, we discussed the case of a 2-Higgs-Doublet model with
a SM-like Higgs at $125$\,GeV and an additional heavier neutral CP-even
Higgs boson with suppressed couplings to gauge bosons. We demonstrated the
importance of the interference between off-shell contributions of the light
Higgs and the on-shell contribution of the heavy Higgs. If the suppression
of the couplings of the heavy Higgs boson to gauge bosons is not too strong,
the $H\rightarrow VV^{(*)}$ channel can in this way lead to the detection of
a heavy Higgs boson at a linear collider, even beyond the kinematical limit
for producing a pair of heavy Higgs bosons, $H$ and $A$.

\begin{table}[tb]
\begin{center}
\begin{tabular}{| c || c | c || c | c |}
\hline
$\sqrt{s}$ & $\sigma^{Z_1 Z_2 Z_3}_{\text{off}}$ ($\sigma^{ZZZ}_{\text{off}}$) & $\Delta_{\text{off}}^{Z_1 Z_2 Z_3}$ ($\Delta_{\text{off}}^{ZZZ}$)
& $\sigma^{\nu \bar{\nu}ZZ}_{\text{off}}$ & $\Delta_{\text{off}}^{\nu \bar{\nu} ZZ}$
\\\hline\hline
$250$\,GeV & $3.12(3.12)$\,fb   & $0.03(0.03)$\,\% & $0.490$\,fb & $0.12$\,\%\\\hline
$300$\,GeV & $2.36(2.40)$\,fb   & $0.46(1.83)$\,\% & $1.12$\,fb & $0.40$\,\%\\\hline
$350$\,GeV & $1.71(1.82)$\,fb   & $1.82(7.77)$\,\% & $1.91$\,fb & $0.88$\,\%\\\hline
$500$\,GeV & $0.802(0.981)$\,fb & $7.20(24.1)$\,\% & $4.78$\,fb & $2.96$\,\%\\\hline
$1$\,TeV   & $0.242(0.341)$\,fb & $30.9(50.9)$\,\% & $15.0$\,fb & $13.0$\,\%\\\hline
  \hline
$\sqrt{s}$ & $\sigma^{ZWW}_{\text{off}}$ & $\Delta_{\text{off}}^{ZWW}$ 
& $\sigma^{\nu \bar{\nu} WW}_{\text{off}}$ & $\Delta_{\text{off}}^{\nu \bar{\nu} WW}$
\\\hline\hline
$250$\,GeV & $76.3$\,fb & $0.03$\,\% & $3.98(3.99)$\,fb & $0.13(0.12)$\,\%\\\hline
$300$\,GeV & $57.7$\,fb & $0.42$\,\% & $9.07(9.08)$\,fb & $0.29(0.26)$\,\%\\\hline
$350$\,GeV & $41.4$\,fb & $0.92$\,\% & $15.5(15.5)$\,fb & $0.49(0.43)$\,\%\\\hline
$500$\,GeV & $18.6$\,fb & $2.61$\,\% & $38.2(38.1)$\,fb & $1.21(0.96)$\,\%\\\hline
$1$\,TeV   & $4.58$\,fb & $11.0$\,\% & $110.8(108.9)$\,fb & $4.45(2.78)$\,\%\\\hline
\end{tabular}
\end{center}
\vspace{-5mm}
\caption{Inclusive cross sections $\sigma_{\text{off}}(0,\sqrt{s}-m_Z)$
for $e^+e^-\rightarrow ZH\rightarrow ZVV$
and $\sigma_{\text{off}}(0,\sqrt{s})$ for $e^+e^-\rightarrow \nu\bar{\nu} H \rightarrow \nu\bar{\nu} VV$
for $P(e^+,e^-)=(0.3,-0.8)$ and
relative size of the off-shell contributions $\Delta_{\text{off}}$ in \%.
In brackets we add the results averaging over the $ZZ$ pairs
for $e^+e^-\rightarrow ZZZ$ and taking into account the $t$-channel Higgs contribution for
$e^+e^-\rightarrow \nu\bar{\nu} WW$. $\Delta_{\text{off}}$ is independent
of the polarisation. Table taken from Ref.~\cite{Liebler:2015aka}.}
\label{table:higgs-gmp}
\end{table}

\section{Top quark and electroweak physics in the SM and beyond} 
Precision measurements of the properties of the top quark and the electroweak gauge bosons present an important part of the physics program of future $e^+e^-$ colliders as they offer additional opportunities to reveal signs of physics beyond the SM and provide important input to the global interpretation of Higgs properties. 
A crucial role plays the measurement of the top quark mass in continuums measurements as well as via threshold scans, these studies are covered in the project B11 of this SFB~\cite{Bach:2017ggt,Reuter:2018ubq}.
The impact of measurements at the ILC and in particular of its $Z$-pole option on
the global electroweak fit have been studied for instance in the  project B8 of this SFB~\cite{Baak:2014ora}.

\subsection{Off-shell processes in top quark pair production}
\begin{figure}[tb]
\centering
  \begin{subfigure}{.485\linewidth}
    \includegraphics[width=\textwidth]{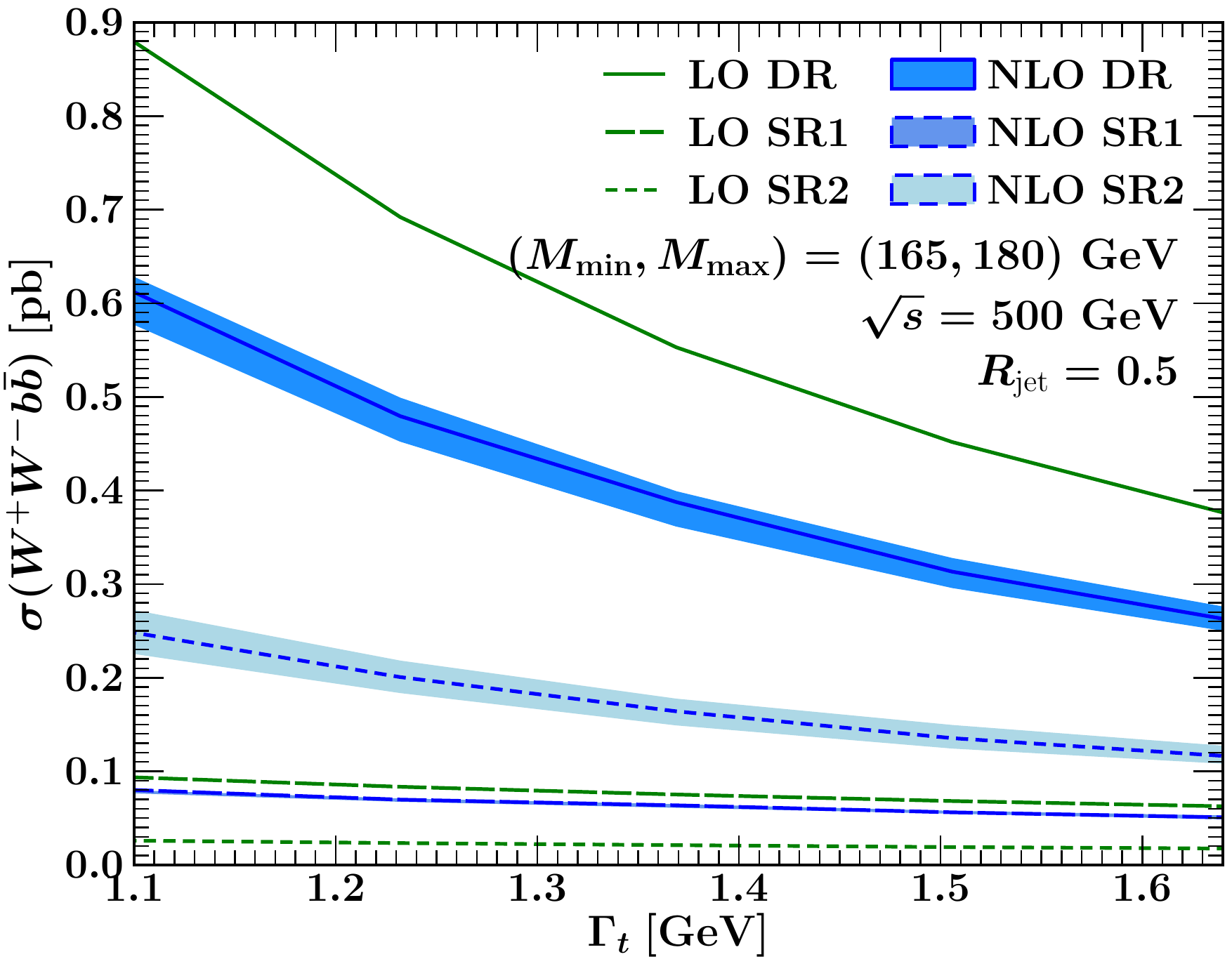}
\caption{}  
  \end{subfigure}
	\hfill
  \begin{subfigure}{.485\linewidth}
    \includegraphics[width=\textwidth]{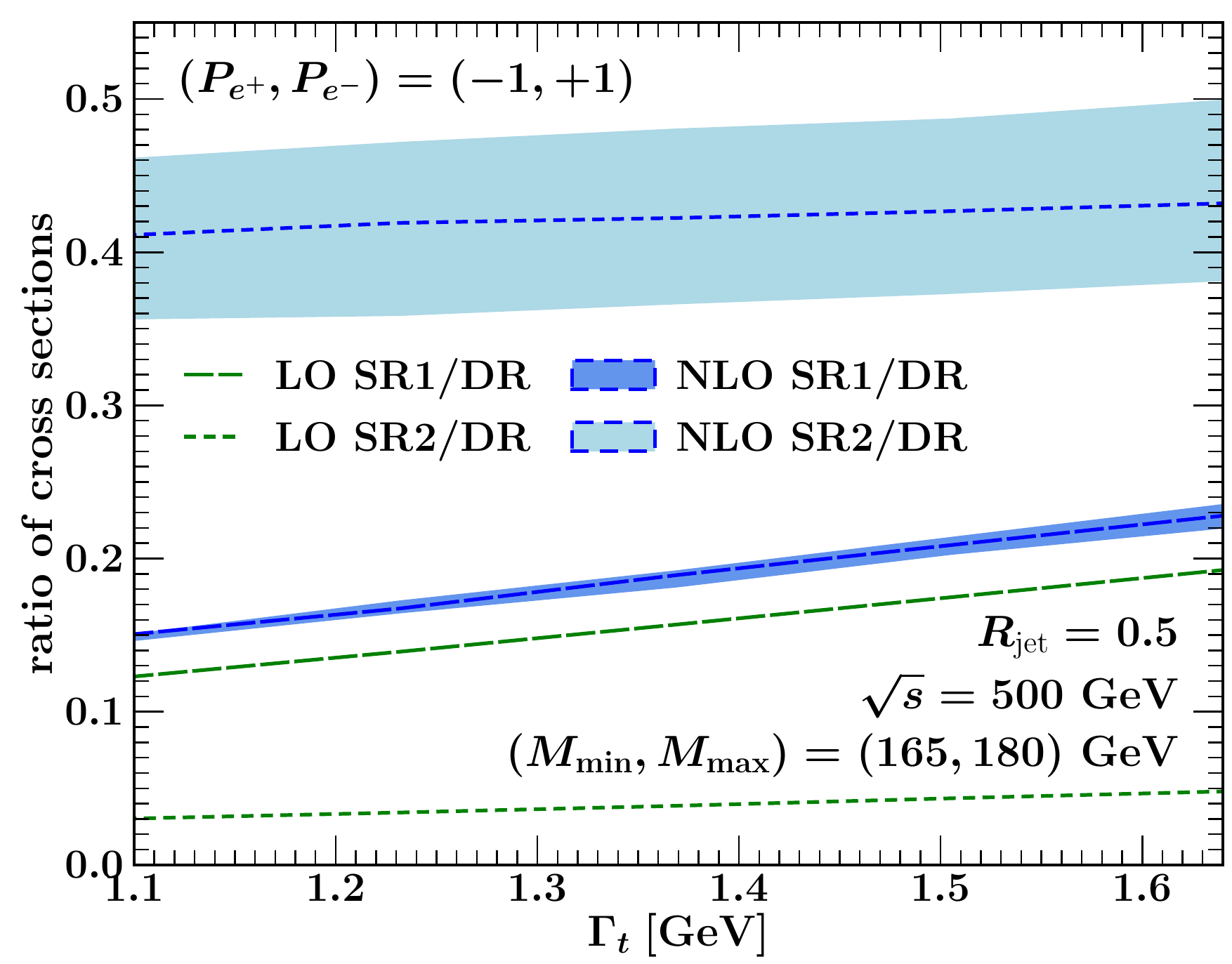}
  \caption{}
  \end{subfigure}
  \caption{
(a) Dependence of cross section $\sigma(e^+e^-\to W^+W^- b \bar{b})$/[pb] at $\sqrt{s}=500$~GeV
at LO and NLO on the top-quark width. `DR' denotes the double resonance region (= two intermediate top-quarks), 
$SR1$ and $SR2$ (= one intermediate top-quark) or the choice $R_{\rm jet} = 0,5$ and mass region $M_{\rm min} = 165$~GeV and $M_{\rm min} = 180$~GeV.
(b) Dependence of the ratios of the single resonant cross sections (SR1 = long dashed,  SR2 = short dashed) to the double resonant 
cross section (DR) on $\Gamma_t$ and for
polarized beams $(P_{e^+}, P_{e^-}) =(-1, +1)$ at $\sqrt{s} = 500$~GeV
(LO = green lines, NLO = blue regions). Figures taken from Ref.~\cite{Liebler:2015ipp}.
}
  \label{fig:top-gmp}
\end{figure}

Within this project, the prospocts to access the top-quark width 
by exploiting off-shell regions in the process $e^+e^-\rightarrow W^+W^-b\bar{b}$ have been studied~\cite{Liebler:2015ipp}.
Next-to-leading order QCD corrections have been taken into account and we showed 
that carefully selected ratios of off-shell regions to on-shell regions 
in the reconstructed top and anti-top invariant mass spectra are, \emph{independently} 
of the coupling $g_{tbW}$, but sensitive to the top-quark width. 
We have examined the structure of reconstructed top-quark
masses allowing for a detailed understanding of the double-, single- and non-resonant contributions of the 
total cross section. The ratio of single-resonant to 
double-resonant cross section contributions, cf. Fig.~\ref{fig:top-gmp}, 
is sensitive to the top-quark width whilst simultaneously
being independent of the $g_{tbW}$ coupling.   The central results of \cite{Liebler:2015ipp} are the in-depth investigation of this ratio. 
We have shown that with a careful choice of the single-resonant
region of the cross section, such a ratio can successfully be 
exploited to extract the width at an $e^+e^-$ collider. We have explored the effects that variations
in both the jet radius as well as the resonance window (in which reconstructed top quarks are defined to be resonant)
have on the ratios. We find that 
attainable accuracies of $< 200$~MeV for determining $\Gamma_t$ are already possible with unpolarised beams at $\sqrt{s}=500$~GeV.
Using polarised beams or higher centre of mass energies would lead to
an enhanced sensitivity to $\Gamma_t$. We note that this is comparable to the accuracies quoted in the literature obtained from 
invariant-mass lineshape fitting, but that the results can significantly be improved by further exploiting this methods including polarised 
initial states. 


\subsection{Electroweak precision measurements}

As an early synergy within the SFB, the fermionic electroweak two-loop corrections to $sin^2 \theta_{eff}^{b \bar{b}}$ where calculated in close collaboration with the project B4~\cite{Awramik:2008gi}. An accurate theoretical prediction is indispensable for the interpretation of $b$ quark asymmetry measurements --- at the $Z$ pole, or at the 250\,GeV stage of the ILC. It was found that these corrections were sizable, especially for the by now known value of the Higgs mass of about 125\,GeV. The experimental capabilities of the ILC at 250\,GeV for measuring the couplings of the $b$ quark to the $Z$ boson have been studied recently by our Paris colleagues~\cite{Bilokin:2017lco}, showing that considerable improvements w.r.t.\ LEP can be reached even at 250\,GeV.

Another important measurement at future $e^+e^-$ colliders will be triple gauge couplings, especially those involving $W$ bosons. The relevant processes like $W$ pair production or single-$W$ production are highly sensitive to the beam polarisations. Therefore it is important to understand a) the role of beam polarisation in disentangling the effects of the various possible anomalous couplings and b) whether the triple gauge couplings and the actual value of the luminosity-weighted average polarisation cn be extracted simultaneously from the data. The ILC prospects have been studied in the SFB-B1 project w.r.t.\ both aspects at all relevant center-of-mass energies ~\cite{marchesini11:_tgcs_and_pol_at_ILC_and_leakage_in_calo, Rosca:2016hcq, Karl:2017let}. 

\begin{figure}[tb]
\centering
  \begin{subfigure}{.485\linewidth}
    \includegraphics[width=\textwidth]{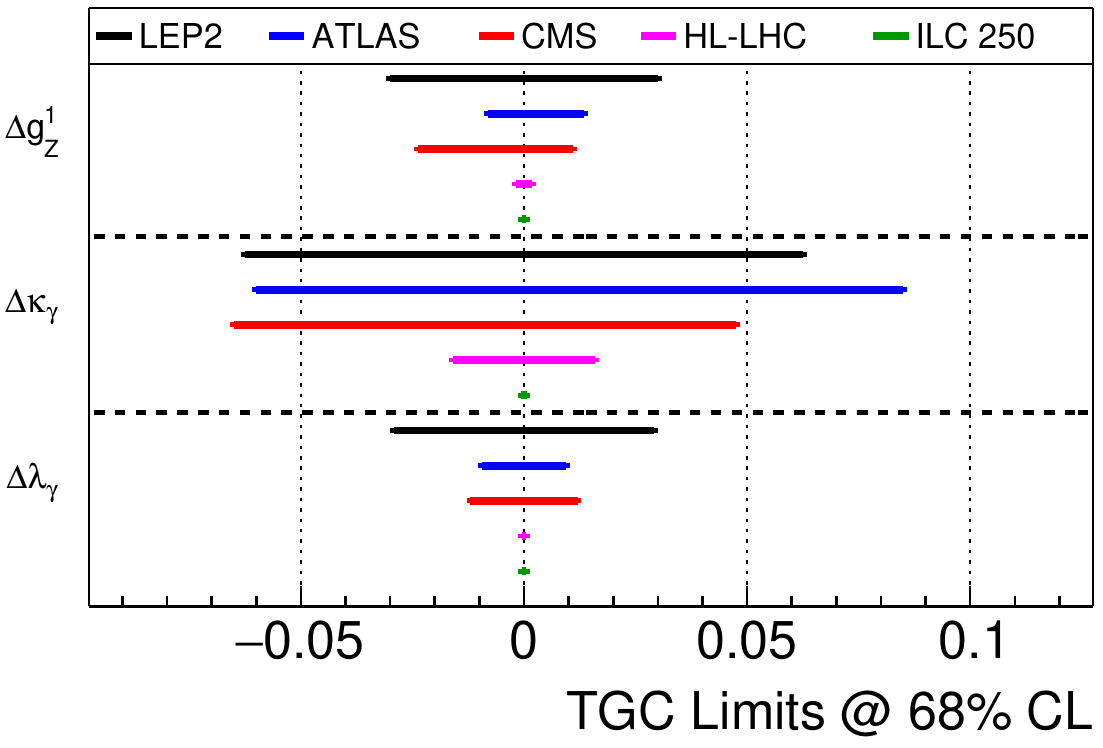}
    \caption{}
    \label{fig:TGCsingle}
  \end{subfigure}
	\hfill
  \begin{subfigure}{.485\linewidth}
    \includegraphics[width=\textwidth]{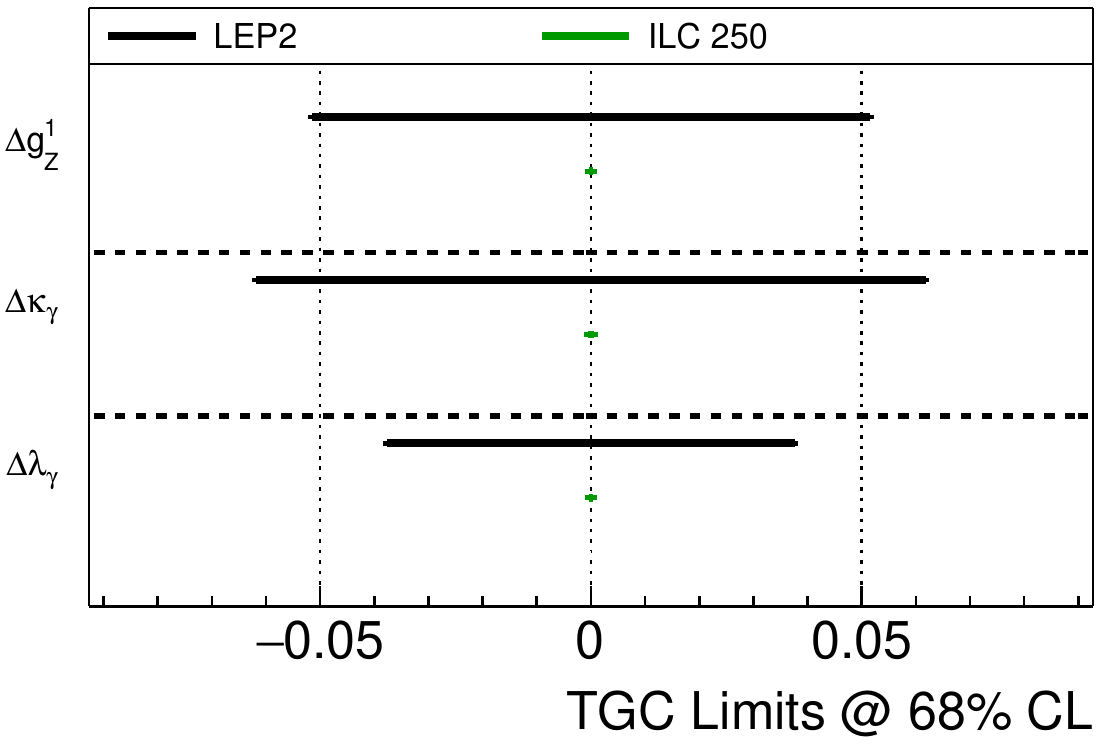}
    \caption{}
    \label{fig:TGCmulti}
  \end{subfigure}
  \caption{Achieved and achievable precisions on anomalous charged triple gauge couplings at LEP, (HL-)LHC and the ILC. The ATLAS and CMS results correspond to their respective analysis basedon the 8\,TeV data. (a) from single-parameter fits (b) from the simultaneous fit of all three couplings, is so far not considered feasible at the LHC. Figures taken from Ref.~\cite{Fujii:2017vwa}.}
  \label{fig:tgc}
\end{figure}

Figure~\ref{fig:tgc} compares the most recent study of the ILC prospects at a first 250\,GeV stage in comparison to the final LEP2 results, the current ATLAS and CMS measurements based on their 8\,TeV data as well as HL-LHC projections~\cite{Karl:2017let}. Thereby Fig.~\ref{fig:TGCsingle} shows the achieved or expected uncertainties when only a single parameter at the time is allowed to differ from its SM value, while Fig.~\ref{fig:TGCmulti} is based on a simulataneous extraction of all three considered couplings, which to date is not considered feasible from hadron collider measurements. These projections are included in the EFT-based interpretation of Higgs measurements discussed in the previous section.


\section{Supersymmetry} 
Over the course of the SFB676, the perspective on searches for direct production of new particles and on the determination of their properties changed drastically: In the earlier phases the work was focussed on the prospects for precision measurements on particles which were assumed to be discovered soon at the LHC but the determination of the specific properties and the distinction of the models was expected to be covered by the LC. With the absence of early discoveries beyond the Higgs boson, the question whether an $e^+e^-$ collider could still discover new particles became of higher and higher relevance. This lead e.g.\ in 2012 to a comprehensive review of the impact of LHC 8\,TeV results and other constraints on the MSSM parameter space and to the definition of new SUSY benchmarks for Linear Collider studies~\cite{Baer:2013ula}. A recent summary of the BSM opportunities at the ILC, prepared with leading contributions from the SFB-B1 project, can be found in~\cite{List:2017jck}. 
For an comprehensive  review on physics at a LC, providing also on overview about theoretical frameworks, see \cite{Moortgat-Pick:2013awa,Moortgat-Picka:2015yla}.
In the following, we will highlight a few individual results obtained with this SFB project.



\subsection{Determination of particle properties}
The challenging task in particle physics ---after the observation of new physics signal--- is the determination of the underlying new physics model. Since the SM is not only highly consistent with all experimental results so far but also from theoretical point of view,
there is no clear direction for a BSM model. Therefore it is even more important to develop strategies how to measure the properties 
of possible new physics candidates precisely and in a model-independent way.
One of the characteristics of several new physics model, as for instance in SUSY or in Universal Extra Dimension (UED) models,  is the spin of the new particles.

\subsubsection{Spin determination}
The spin of supersymmetric particles can be determined unambiguously at $e^+e^-$ colliders. In \cite{Choi:2006mr}, we showed
for a characteristic set of non-colored supersymmetric particles -- smuons, selectrons, and charginos/neutralinos, how to determine
the spin in a model-independent way via three different steps: analyzing the threshold behavior of the excitation curves for pair production in $e^+ e^-$ collisions, the angular distribution in the production process and decay angular distributions, 
For the production of spin-0 sleptons (for selectrons close to threshold),
it turns out that the $\sin^2\theta$-law for the
production is a unique signal of the spin-0 character. However, while the observation of the
$\sin^2\theta$-angular-distribution is sufficient for sleptons, the $\beta^3$
onset of the excitation curve is a necessary but not a sufficient condition for the
spin-0 character. In the case of spin-1/2-particles (chargino/neutralino sector),
neither the onset of excitation curves nor the angular distributions in the
production processes provide unique signals of the spin quantum numbers.
Here, decay angular distributions provide a unique signal for the chargino/neutralino spin $J = 1/2$, albeit
at the expense of more involved experimental analyses, cf.\ Table~\ref{tab:summary}. 

\begin{table}[tb]
\centering
\begin{tabular}{|c||c|cc|cc|}
\hline
\multicolumn{6}{|c|} {Threshold Excitation and Angular Distribution}\\
 \hline \hline
SUSY & particle   & $\tilde{\mu}$  & $\tilde{e}$ & $\tilde{\chi}^\pm$
     & $\tilde{\chi}^0$ \\
{ }  &  spin &   0      &   0   & 1/2
     & 1/2 \\
\cline{2-6}
{ }  & $\sigma_{thr}$ & $\beta^3$      & $\beta^3$   & $\beta$
     & $\beta^3$ \\
{ }  & $\theta$ dep.  & $\sin^2\theta$  &  thr: $\sin^2\theta$
     & thr: isotropic  & thr: $1+\kappa \cos^2\theta$ \\ \hline \hline
UED  & particle  & $\mu_1$ & $e_1$  & $W^\pm_1$ & $Z_1$ \\
{ }  & spin &  1/2    &  1/2   &  1        &   1 \\
\cline{2-6}
{ }  & $\sigma_{thr}$ & $\beta$      & $\beta$   & $\beta$
     & $\beta$ \\
{ }  & $\theta$ dep.  & $1+\kappa^2\cos^2\theta$  &  thr: isotropic
     & thr: isotropic  & thr: isotropic \\ \hline \hline
General  & particle  & $B[s]$ & $B[s,t,u]$  & $F_{D,M}[s]$ & $F_{D,M}[s,t,u]$ \\
{ }  & spin &  $\geq 1$    &  $\geq 1$   &  $\geq 1/2$  & $\geq 1/2$ \\
\cline{2-6}
{ }  & $\sigma_{thr}$ & $\beta^3$     & $\beta$   & $\beta,\beta^3$
     & $\beta,\beta^3$ \\
{ }  & $\theta$ dep.  & $1+\kappa\cos^2\theta$  &  thr: isotropic
     & $1+\kappa \cos^2\theta$  & thr: $1+\kappa \cos^2\theta$\\ \hline
\end{tabular}
\caption{\label{tab:summary} { The table shows
the general characteristics of spin-$J$ particles, the corresponding
 threshold behavior and the
angular distribution in, for instance, SUSY and UED particle pair production processes.
B and F$_{D,M}$ generically denote bosons and Dirac,
Majorana fermions;
The parameters $\kappa$ [$\kappa \neq -1$]
depend on mass ratios and particle velocities $\beta$. Measurements
of the polar angle distribution in the slepton sector provide 
unique spin-0 assignments. However, for spin-1/2 particles neither threshold
excitation nor angular distributions are sufficient, i.e.\ also a  final state analyses must be performed to determine the 
quantum numbers. Table reprinted from Ref.~\cite{Choi:2006mr} with kind permission of The European Physical Journal (EPJ).}}
\end{table}

\subsubsection{Structure of couplings}
\begin{figure}[tb]
\centering
  \begin{subfigure}{.32\linewidth}
    \includegraphics[width=\textwidth]{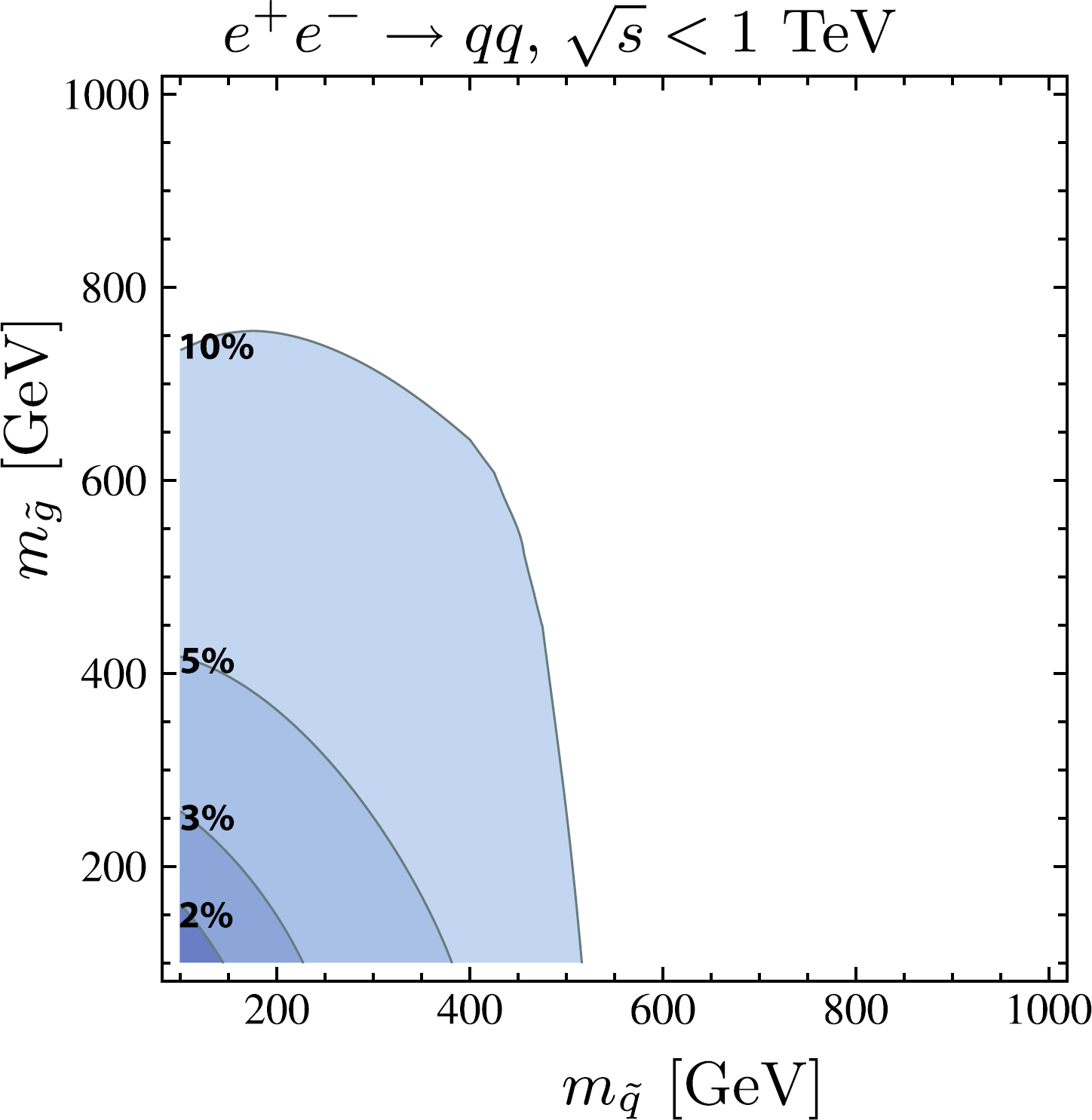}
    \caption{}
  \end{subfigure}
	\hfill
  \begin{subfigure}{.32\linewidth}
    \includegraphics[width=\textwidth]{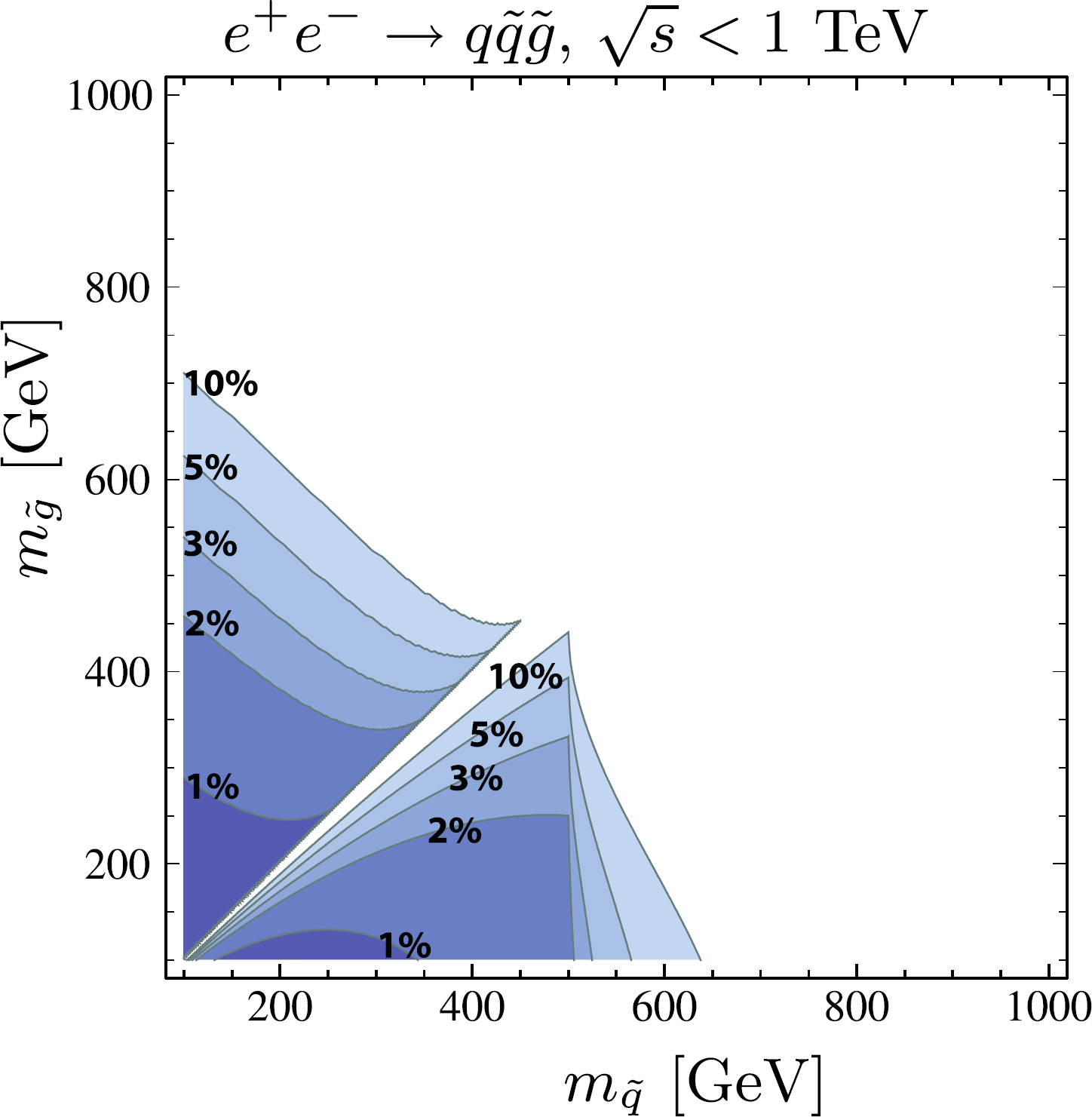}
    \caption{}
  \end{subfigure}
	\hfill
  \begin{subfigure}{.32\linewidth}
    \includegraphics[width=\textwidth]{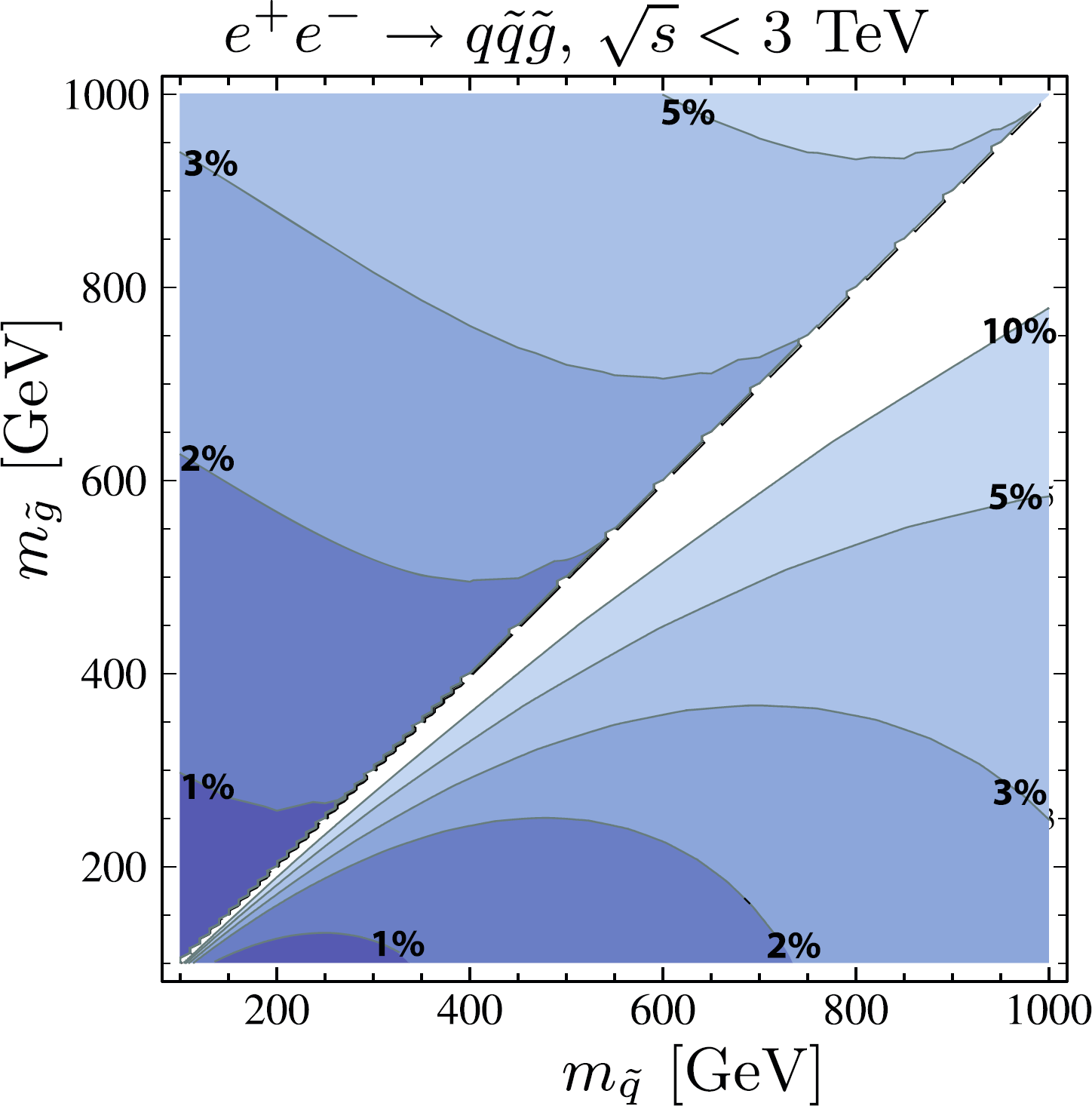}
    \caption{}
  \end{subfigure}
  \caption{  
   Contour plots of the statistical 1-$\sigma$ errors of the Yukawa coupling $\Delta \alpha_s/\alpha_s$ in the 
   $[m_{\tilde{q}},m_{\tilde{g}}]$ mass plane; 
(a)   the indirect channel $e^+e^-\to q q$ at $\sqrt{s}\le 1$~TeV; 
the golden direct channel 
$e^+ e^- \to q \tilde{q} \tilde{g}$, corresponding to 
 (b) a maximal ILC $\sqrt{s}\le 1$~TeV, and (c) up to $\sqrt{s}\le 3$~TeV for CLIC, respectively.
Figures reprinted from Ref.~\cite{Brandenburg:2008gd} with kind permission of The European Physical Journal (EPJ).}
  \label{fig:yukawa-gmp}
\end{figure}

A specific feature for Supersymmetric models is that the coupling characterics are preserved under SUSY transformations.
In order to prove supersymmetry, it is therefore
necessary to verify this feature. For instance, the SUSY Yukawa-couplings have to be proven to be
identical to the corresponding gauge couplings. In the electroweak sector, it has been shown in \cite{Choi:2001ww}, that
the measurements of polarized cross section serve perfectly well for this purpose. In  \cite{Brandenburg:2008gd}, the study has been extended to the coloured sector and it has been examined whether the quark-squark-gluino Yukawa couplings,
 can be determined, complementary to LHC analyses, by 
studying  $q-\tilde{q}-\tilde{g}$ and $\tilde{q}-\tilde{q}-g$ and comparing it with the radiation process  $qqg$
 at a TeV $e^+ e^-$ collider. SUSY QCD corrections at NLO have been included.
 These channels have been investigated to test this fundamental identity between the couplings.
While the golden channel measures the  $q-\tilde{q}-\tilde{g}$ Yukawa coupling, the radiation processes 
$\tilde{q}-\tilde{q}-g$  and $qqg$ determine the
QCD gauge coupling in the $\tilde{q}$-sector and the standard $q$-sector for comparison. A few percent variation in the Yukawa couplings is observable at $e^+e^-$ colliders, depending on the masses of the squark and gluinos, cf. Figs.\ref{fig:yukawa-gmp}. Such a potential is fully complementary to the respecting potential at the LHC where the production of $\tilde{q}$-pairs in $qq$ collisions 
provides a stage for the measuriing the SUSY-QCD Yukawa coupling, requiring
however, an ensemble of auxiliary measurements of decay branching ratios.

\subsubsection{CP properties}
\begin{figure}[tb]
\centering
  \begin{subfigure}{.48\linewidth}
    \includegraphics[width=\textwidth]{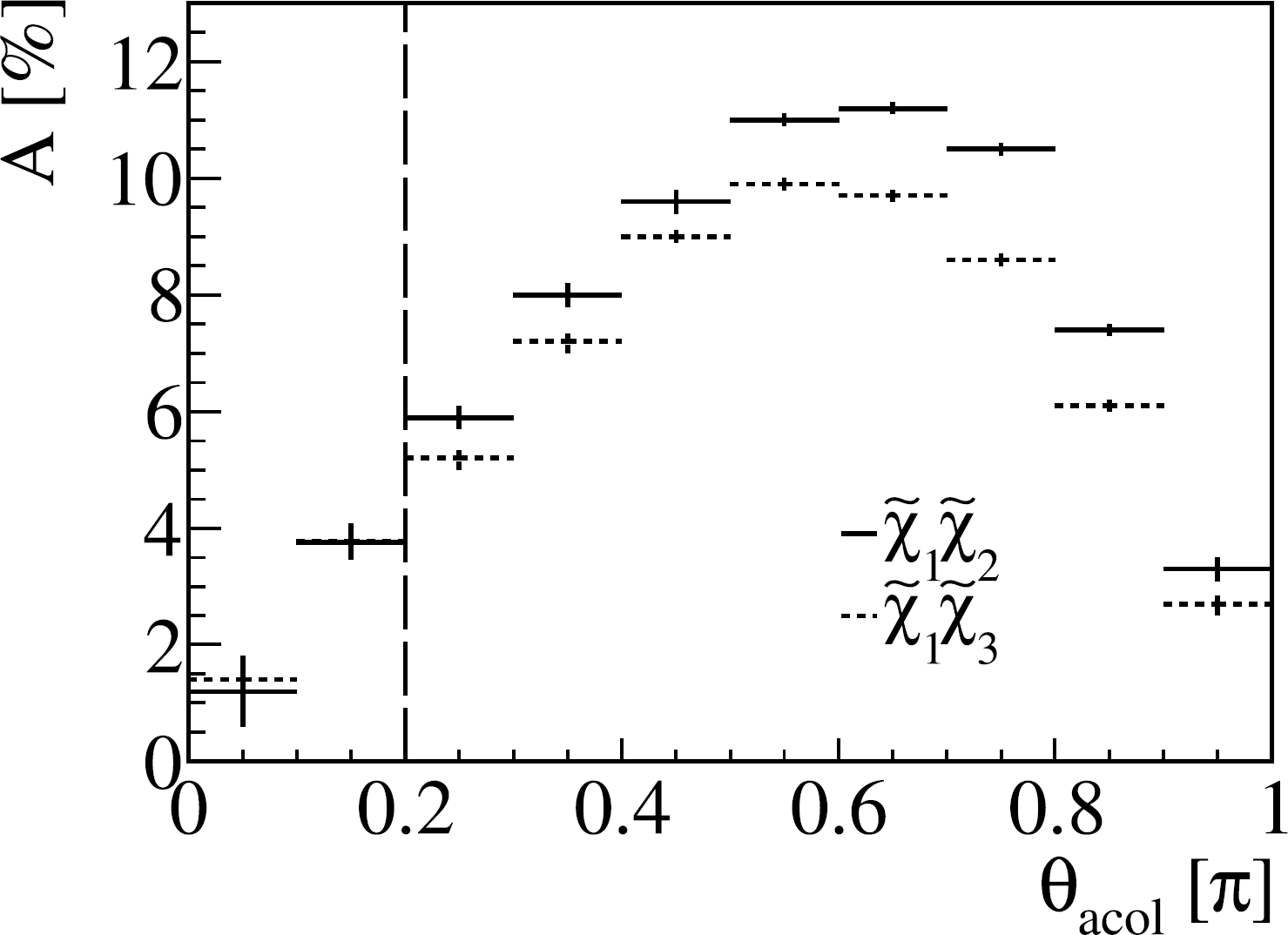}
    \caption{}
  \end{subfigure}
	\hfill
  \begin{subfigure}{.51\linewidth}
   \includegraphics[width=\textwidth]{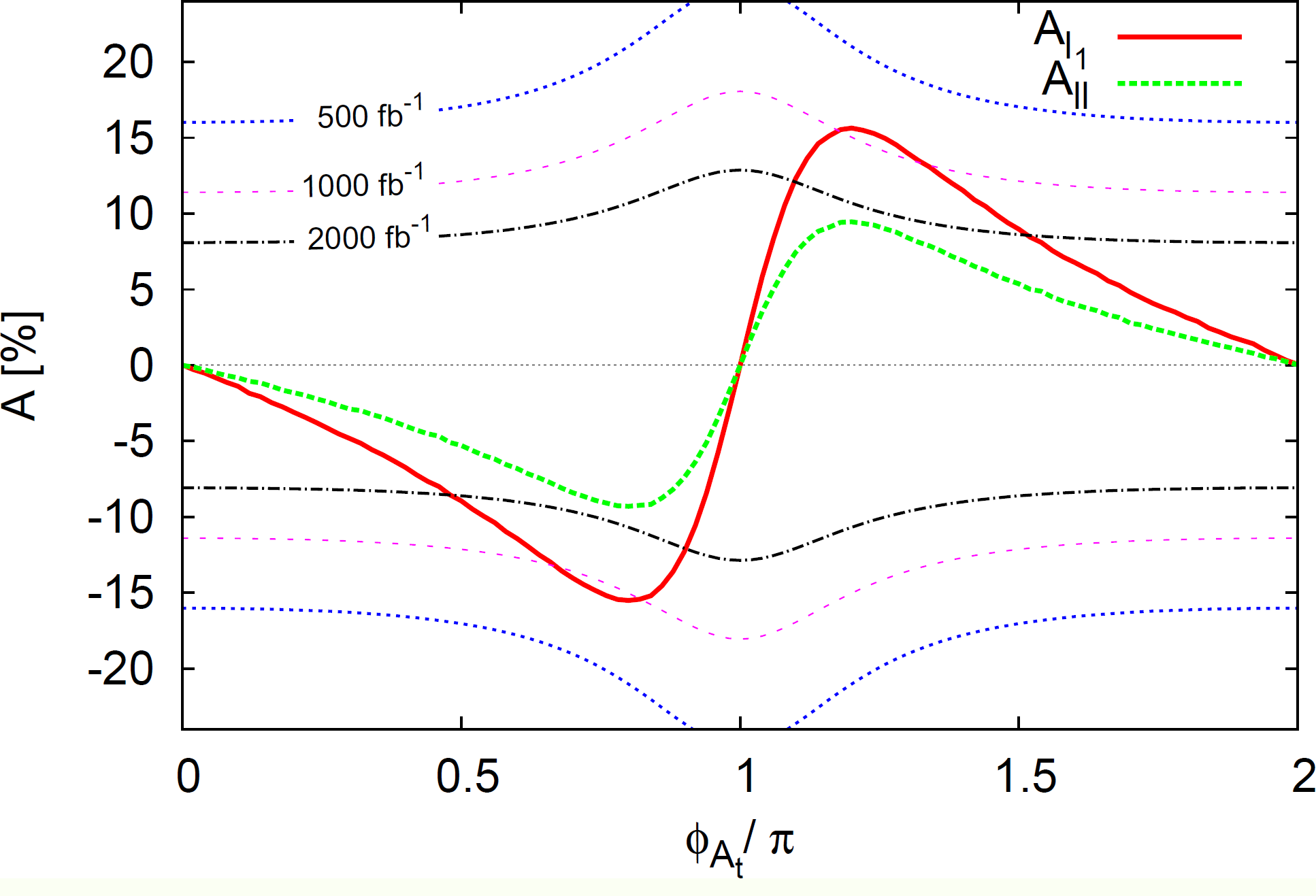}
   \caption{}
  \end{subfigure}
\caption{
(a) The $\theta_{\rm aco}$ dependence of the asymmetries of
$A\left[\vec{p}_{e-},\vec{p}_{\ell-N},\vec{p}_{\ell+F}  \right]_{\tilde{\chi}^0_1\tilde{\chi}^0_2}$ (solid)
and 
$A\left[\vec{p}_{e-},\vec{p}_{\ell-N},\vec{p}_{\ell+F}    \right]_{\tilde{\chi}^0_1 \tilde{\chi}^0_3}$
(dashed).
The cut value used in our analysis is indicated by the dashed line. In each case $10^7$
events were generated and no detector effects are included.
Figure reprinted from Ref.~\cite{Kittel:2011rk} with kind permission of The European Physical Journal (EPJ).
(b) The asymmetries, originating in stop decays into neutralinos, $A_{l_1}$ and $A_{ll}$, along
with the lines showing the asymmetry required for
a 3$\sigma$ observation at a given integrated luminosity of
500~fb$^{-1}$, 
1000~fb$^{-1}$ 
and 2000 fb$^{-1}$ at $\sqrt{s}=1$~TeV in
the case of momentum reconstruction.
Figure taken from Ref.~\cite{Salimkhani:2012dqa}. }
  \label{fig:cpasy-gmp}
\end{figure}

Since SUSY offers naturally new sources  for CP-violation, required for explaining the baryon-antibaryon asymmetry in our Universe, 
it is of high importance to work out to which extent these phases could be determined in experiments. 
The sizes of these phases are constrained by experimental bounds from the electric
dipole moments (EDMs). Such experimental limits generally restrict the CP phases
to be small, in particular the phase $\Phi_\mu$. Cancellations among different contributions to the EDMs can occur so that 
still large CP phases could happen, causing CP-violating signals at colliders. Thus, direct measurements of SUSY CP-sensitive observables are necessary to determine or constrain the phases independently of EDM measurements. The phases
change SUSY particle masses, their cross sections, branching ratios. However, although such CP-even observables
are sensitive to the CP phases, CP-odd (T-odd) observables have to be
measured for a direct evidence of CP violation. In~\cite{Kittel:2011rk,Terwort:2012sy} triple product asymmetries
have been studied in an interdisciplinary theo-exp endeavour resulting  a first experimentally-oriented analysis based on a full detector
simulation with regard to the observation of CP asymmetries, cf. Fig.\ref{fig:cpasy-gmp} (a): the process $e^+e^- \to \tilde{\chi}^0_i \tilde{\chi}^0_j$ and subsequent leptonic two-body decays 
$\tilde{\chi}^0_i \to \tilde{l}_R l, \tilde{l}_R \to \tilde{\chi}^0_1 l$, for $l=e, \mu$ has been calculated and the 
expected triple product asymmetry between the incoming $e^-$ and final leptons $\ell^+$, $\ell^-$ has been evaluated, including the relevant Standard Model background processes, a realistic beam energy spectrum as well as  beam backgrounds. Assuming an integrated luminosity of 500~fb$^-1$ and simultaneous beam polarization of $P_{e^-}=+ 80\%$ and $P_{e^+}=-60\%$ a relative measurement accuracy of 10\% for the CP-sensitive asymmetry is achievable.
We demonstrate that our method of signal selection using kinematic reconstruction can be applied  to a broad class of scenarios and it allows disentangling processes with similar kinematic properties. 

In \cite{Salimkhani:2012dqa} another channel for exploiting  CP-odd observables has been studied: triple product correlations originating from $\tilde{t}_1$ decays into neutrlinos $\tilde{\chi}^0_2$:
$T_{l1}^{\mp}=
(\vec{p}_{l1}\cdot (\vec{p}_W^{\mp} \times \vec{p}_t)$ , $
T_{ll}^{\mp}=
(\vec{p}_{b}\cdot (\vec{p}_{l^+} \times \vec{p}_{l^-})$. 
Assuming a successful momentum reconstruction a maximal asymmetry can be observed with at least 1000 fb$^{-1}$ collected data.
The result showed that the CP violating phase $\phi_{A_t}$ of the trilinear top coupling accounted
for a maximal triple product asymmetry of approximately 15.5 \%, cf. Fig.~\ref{fig:cpasy-gmp} (b). Under the assumption of successful
momentum reconstruction, this asymmetry could be measured for 2000 fb$^{-1}$ collected data in the region
of a maximal CP violating angle, $1.10 \pi < \phi_{A_t} < 1.5\pi$. 
With an integrated luminosity of 1000 fb$^{-1}$ the
asymmetry could still be exposed close to the maximum ($1.18\pi  < \phi_{A_t} < 1.33\pi$).
The results show that a future linear collider with high luminosity is essential.

\subsubsection{Majorana character}
SUSY offers not only Dirac-type fermions but also Majorana-type massive fermions (where the particle is its own antiparticle).
This property is particularly difficult to prove uniquely experimentally, it is affected by the impact of spin correlations 
 in production$\times$three-body decays~\cite{MoortgatPick:2002iq}, threshold behaviours~\cite{Choi:2004jf} and two-body 
 decays~\cite{Choi:2003fs}.

\begin{figure}[tb]
\centering
  \begin{subfigure}{.485\linewidth}
    \includegraphics[width=\textwidth]{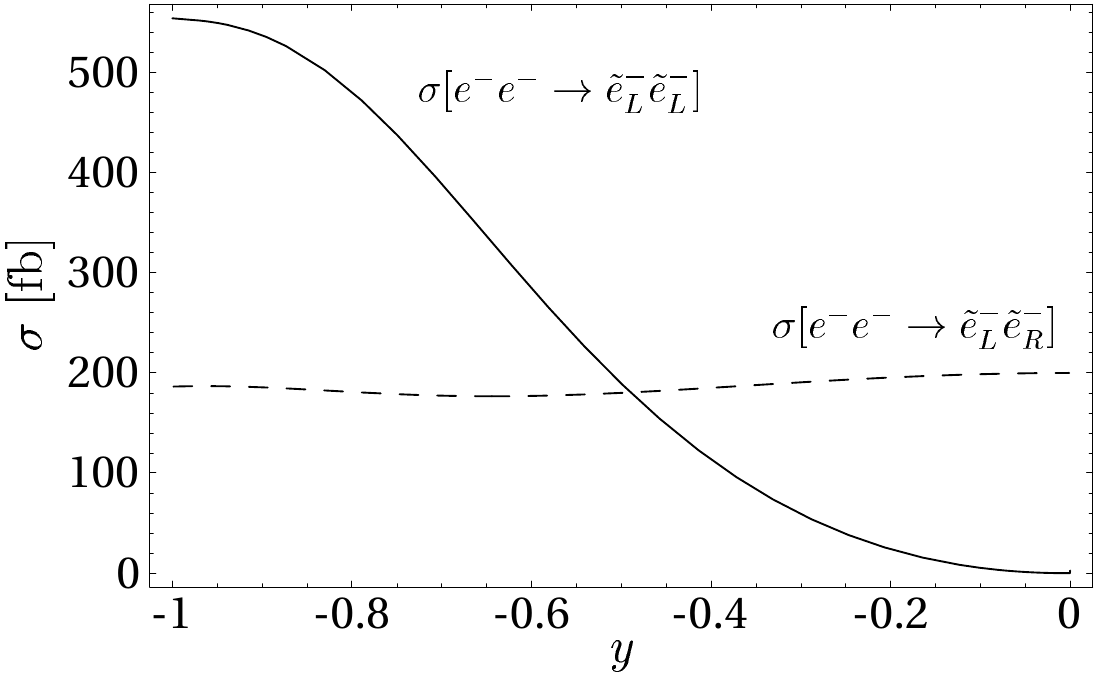}
    \caption{}
  \end{subfigure}
	\hfill
  \begin{subfigure}{.495\linewidth}
    \includegraphics[width=\textwidth]{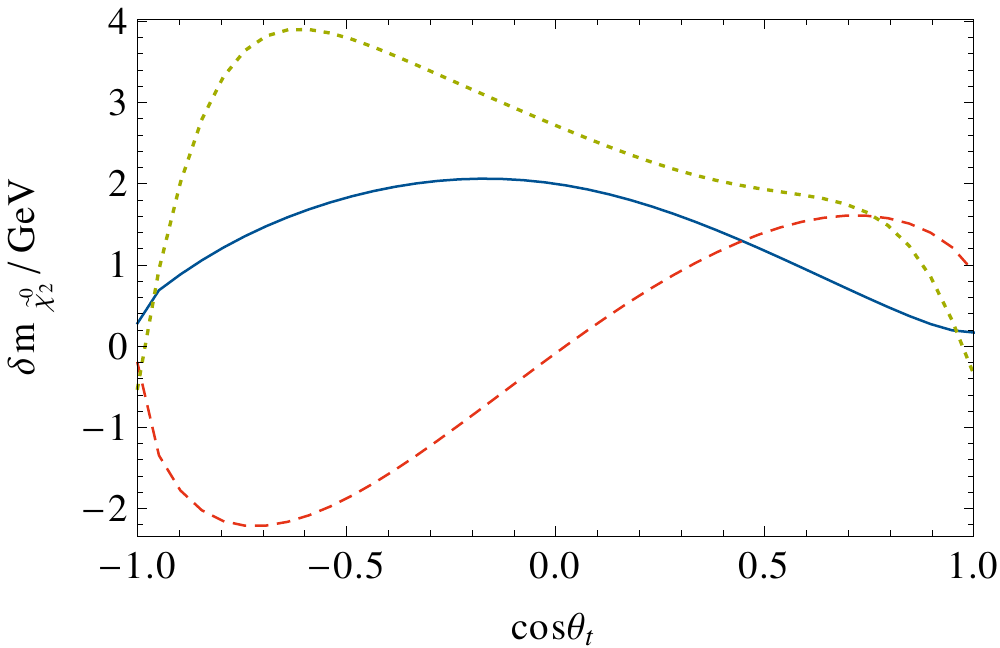}
    \caption{}
  \end{subfigure}
  \caption{(a) Partonic cross sections for same-sign selectron production as a functions of the Dirac/Majorana control parameter y,
for $\sqrt{s} = 500$~GeV and SPS1a' parameters. Not shown is the cross section for 
$e^- e^- \to \tilde{e}^{-}_L \tilde{e}^{-}_R$,
which, apart from the
different normalization, shows a similar behavior as the cross section for 
$e^- e^- \to \tilde{e}^{-}_L \tilde{e}^{-}_L$. 
Reprinted figure with permission from Ref.~\cite{Choi:2008pi}. Copyright (2008) by the American Physical Society.
(b) 
One-loop corrections to the masses of neutralinos $\tilde{\chi}^0_2$
as a function of the stop mixing angle $\cos \theta_t$, for three scenarios S1 ($(M_1, M_2, \mu)=(125,250,180)$~GeV)
(blue), S2 ($(M_1,M_2,\mu)=(125,2000,180)$~GeV) (red, dashed) and S3 ($(M_1,M_2,\mu)=(106,212,180)$~GeV) 
(green, dotted).
Figure reprinted from Ref.~\cite{Bharucha:2012ya} with kind permission of The European Physical Journal (EPJ).
}
  \label{fig:majo-gmp}
\end{figure}

The study ~\cite{Choi:2008pi} addresses  the comparison of the production of Majorana-like neutralinos and gluinos in the MSSM with that of Dirac-like neutralinos and gluinos within the framework of $N=2$ MSSM. Decays of such self-conjugate particles generate charge symmetric ensembles of final states.  
The LC offers a unique possibility to adjust the experimental conditions particularly well to specific needs of the theories: the $^-e^-$-mode would offer unique possibilities to test the Majorana- versus Dirac-exchange-character of the involved processes. 
In the study, it has been analyzed to which extent like-sign dilepton production in the process 
$e^- e^-\to \tilde{e}^-\tilde{e}^-$ is affected by the exchange of either Majorana or Dirac neutralinos, see Fig.~\ref{fig:majo-gmp} (left panel). 
Using polarized beams at an $e^-e^-$-LC,
the Dirac/Majorana character can be studied experimentally.
The 'conclusio generalis' of \cite{Choi:2008pi} is that the Majorana theory can be
discriminated from the Dirac theory using like-sign dilepton events at the level of more than 10$\sigma$.

\subsection{Light higgsinos}

\begin{figure}[tb]
\centering
  \begin{subfigure}{.415\linewidth}
    \includegraphics[width=\textwidth]{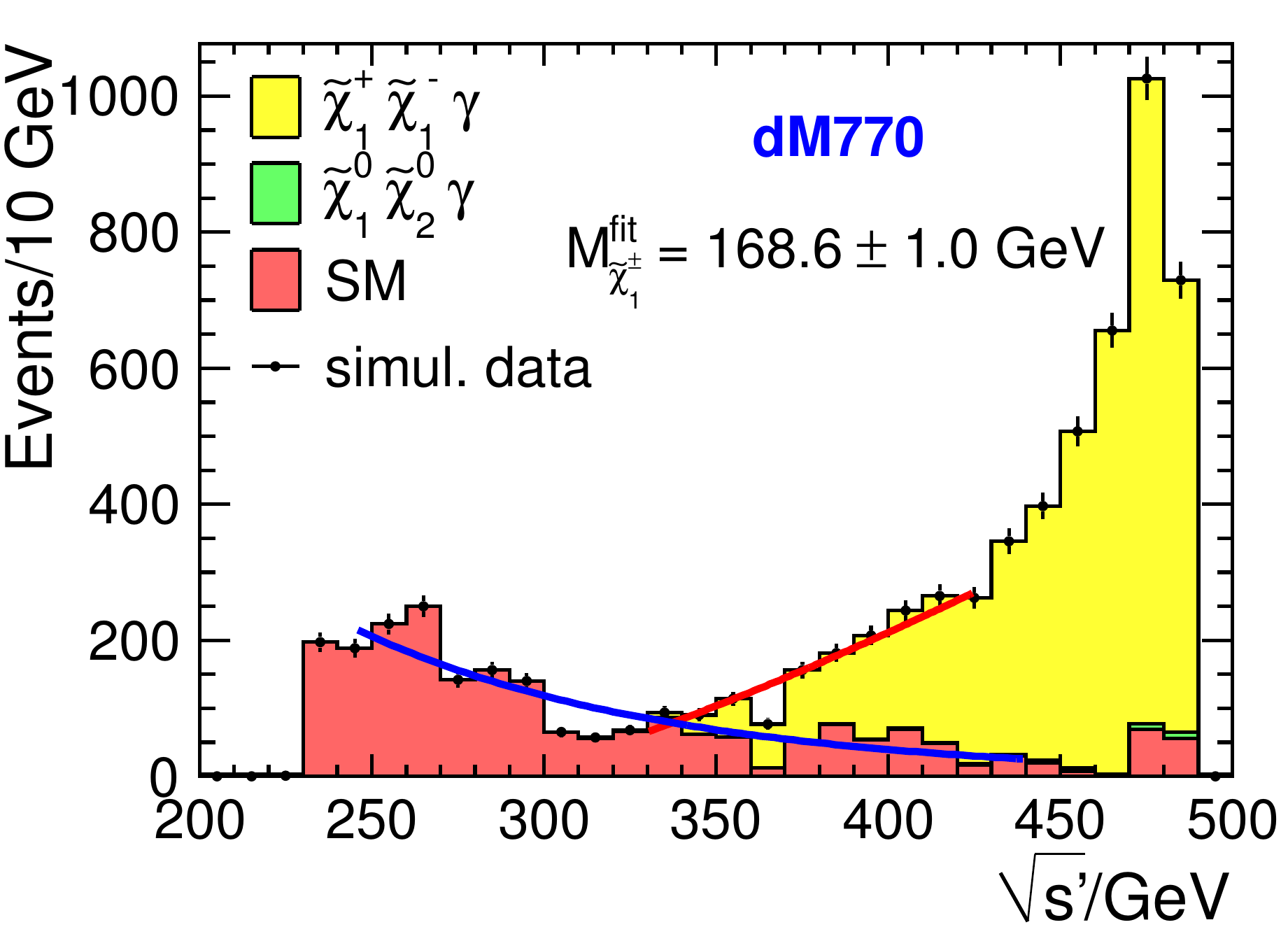}
    \caption{}
    \label{fig:higgsino770_mrec}
  \end{subfigure}
	\hfill
  \begin{subfigure}{.562\linewidth}
    \includegraphics[width=\textwidth]{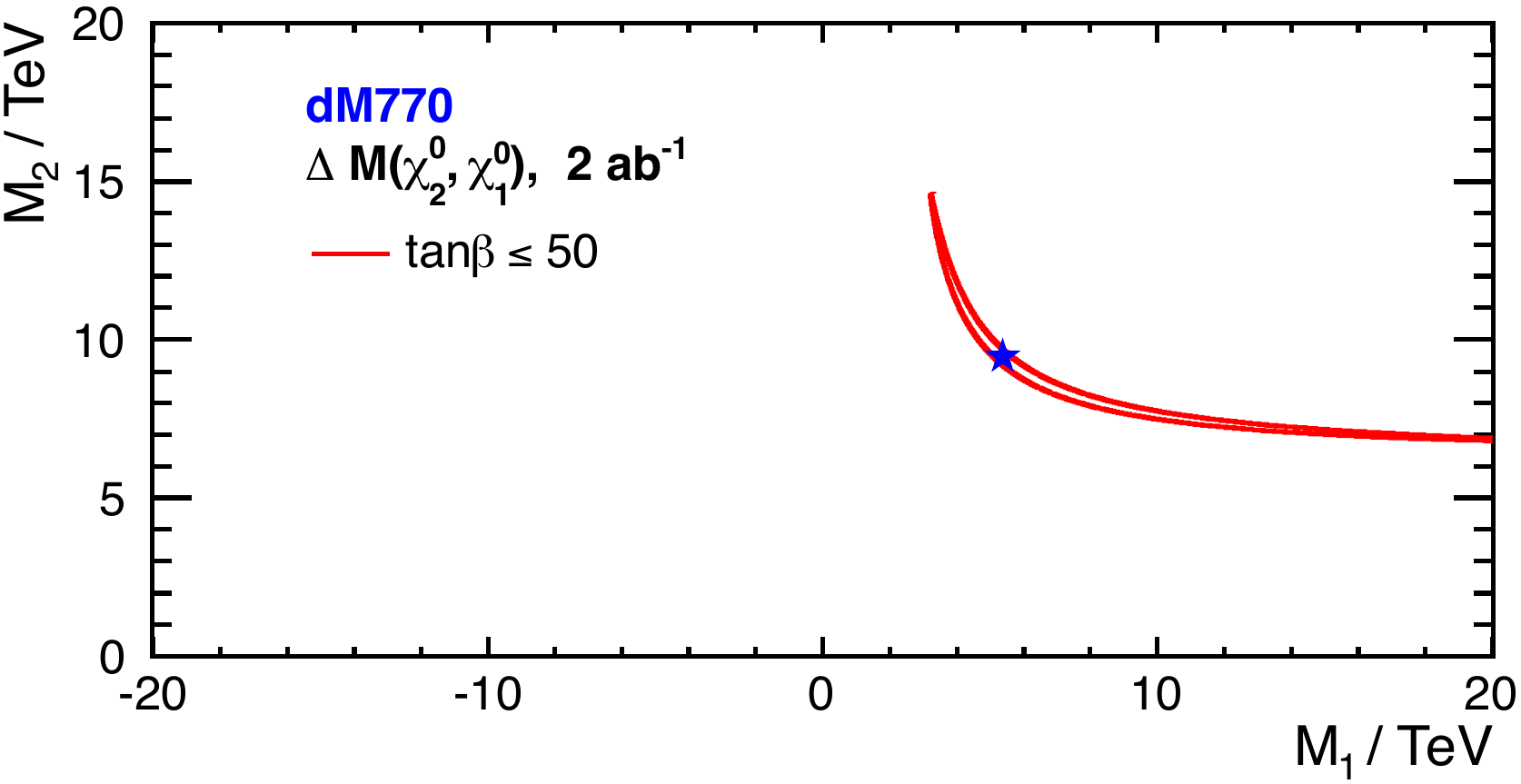}
    \caption{}
    \label{fig:higgsino770_M1M2}
  \end{subfigure}
  \caption{Light higgsinos with sub-GeV mass differences at the ILC: (a) Measurement of the chargino mass from the recoil against an ISR photon for 500\,fb$^{-1}$ at 500\,GeV. (b) Constraint on the multi-TeV gaugino mass parameters $M_1$ and $M_2$ obtained from higgsino mass and cross section measurements at the 500\,GeV ILC, based on an integrated luminosity of 2000\,fb$^{-1}$, split equally between the two opposite-sign beam polarisation states. Figures reprinted from Ref.~\cite{Berggren:2013vfa} with kind permission of The European Physical Journal (EPJ).}
  \label{fig:higgsinos}
\end{figure}

Another important SUSY example are scenarios with light higgsinos. They can be motivated by naturalness arguments~\cite{Baer:2012up}, but also occur in hybrid gauge-gravity mediation models motivated by string theory. The latter case has been studied in close collaboration with the A1 project of this SFB~\cite{Berggren:2013vfa,Sert:293539}. Two benchmark points with higgsino mass splittings of 1.6\,GeV and 770\,MeV, respectively, have been chosen to evaluate the ILC prospects in detailed simulation of the ILD detector concept. This study led to a new awareness within the ILD concept group concerning the importance of designing the detector with sufficient sensitivity to low-momentum particles, and these model-points have become standard benchmarks for the ILD detector optimisation process.
Figure~\ref{fig:higgsino770_mrec} shows the recoil mass of a chargino pair against an ISR photon. From the endpoint of the signal distribution the mass of the chargino can be determined with sub-percent precision already from 500\,fb$^{-1}$. Together with the determination of the mass difference to the LSP from the energy distribution of the visible decay products and the polarised cross sections, the higgsino mass parameter $\mu$ can be determined, and the gaugino mass parameters can be constrained to a narrow region in the multi-TeV regime, which is shown in Fig.~\ref{fig:higgsino770_M1M2}.

In the case of light higgsino scenarios with somewhat larger mass differences of a few GeV, even a full SUSY parameter determination can be carried out. Three different benchmark points with chargino-LSP mass differences of 11, 6 and 2.5\,GeV, respectively have been studied in full detector simulation of the ILD concept, and the resulting percent-level precisions on masses and polarised cross sections have been used as inputs to SUSY parameter fits at the weak scale and at the GUT scale~\cite{Lehtinen:2017vdt, Lehtinen:2018PhD}. Similar as in case of the STC benchmark series discussed above, also in these cases the masses, or, in case of the coloured sector, mass ranges of the unobserved sparticles can be predicted.  The relic density of the LSP can also be well determined, showing in these cases clearly that the LSP provides only a small fraction of the dark matter in the universe, while the rest would need another explanation, like e.g.\ axions --- another important topic in the SFB 676.

\begin{figure}[tb]
\centering
  \begin{subfigure}{.32\linewidth}
    \includegraphics[width=\textwidth]{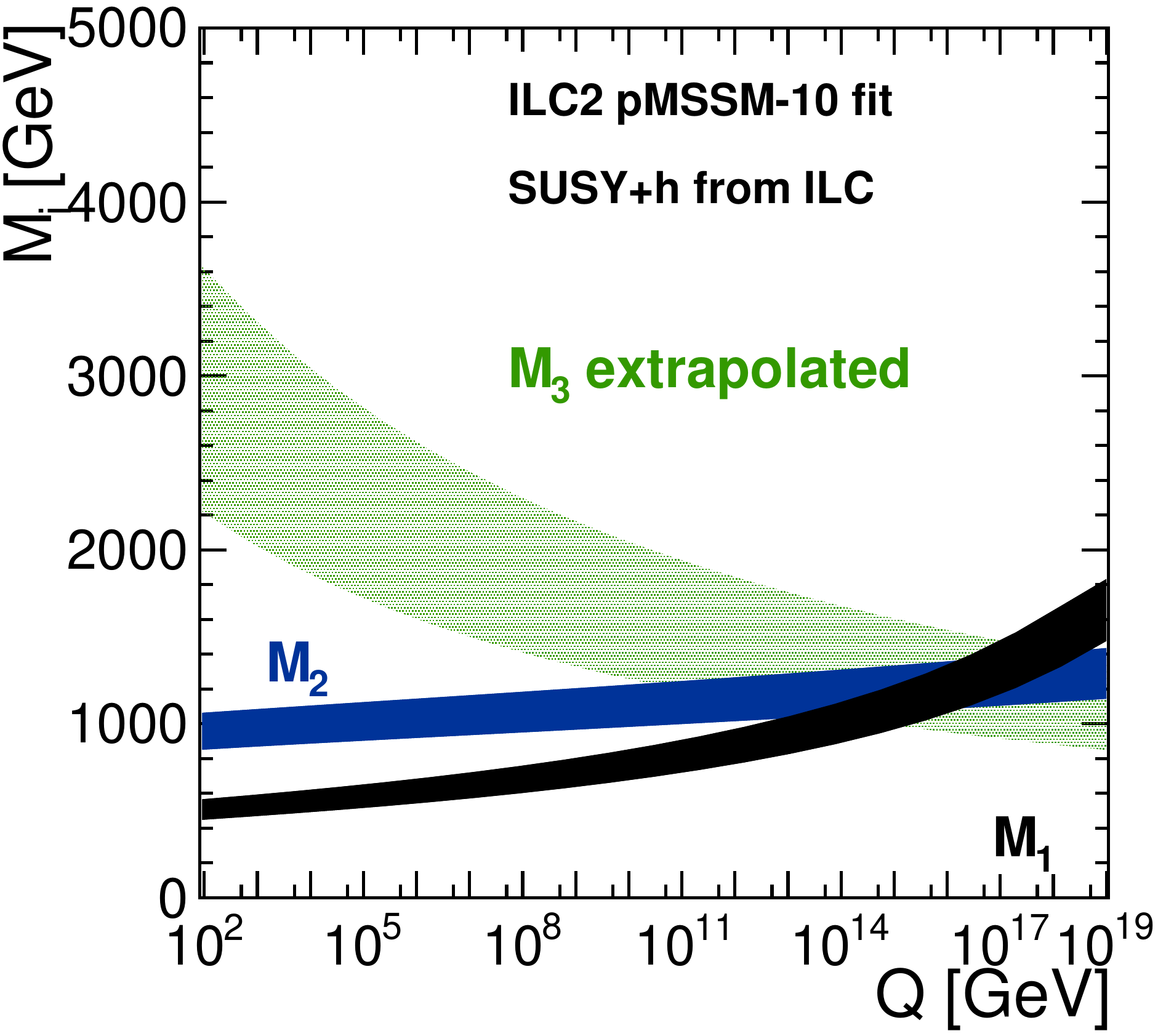}
    \caption{}
    \label{fig:RGE_ILC2}
  \end{subfigure}
	\hfill
  \begin{subfigure}{.32\linewidth}
    \includegraphics[width=\textwidth]{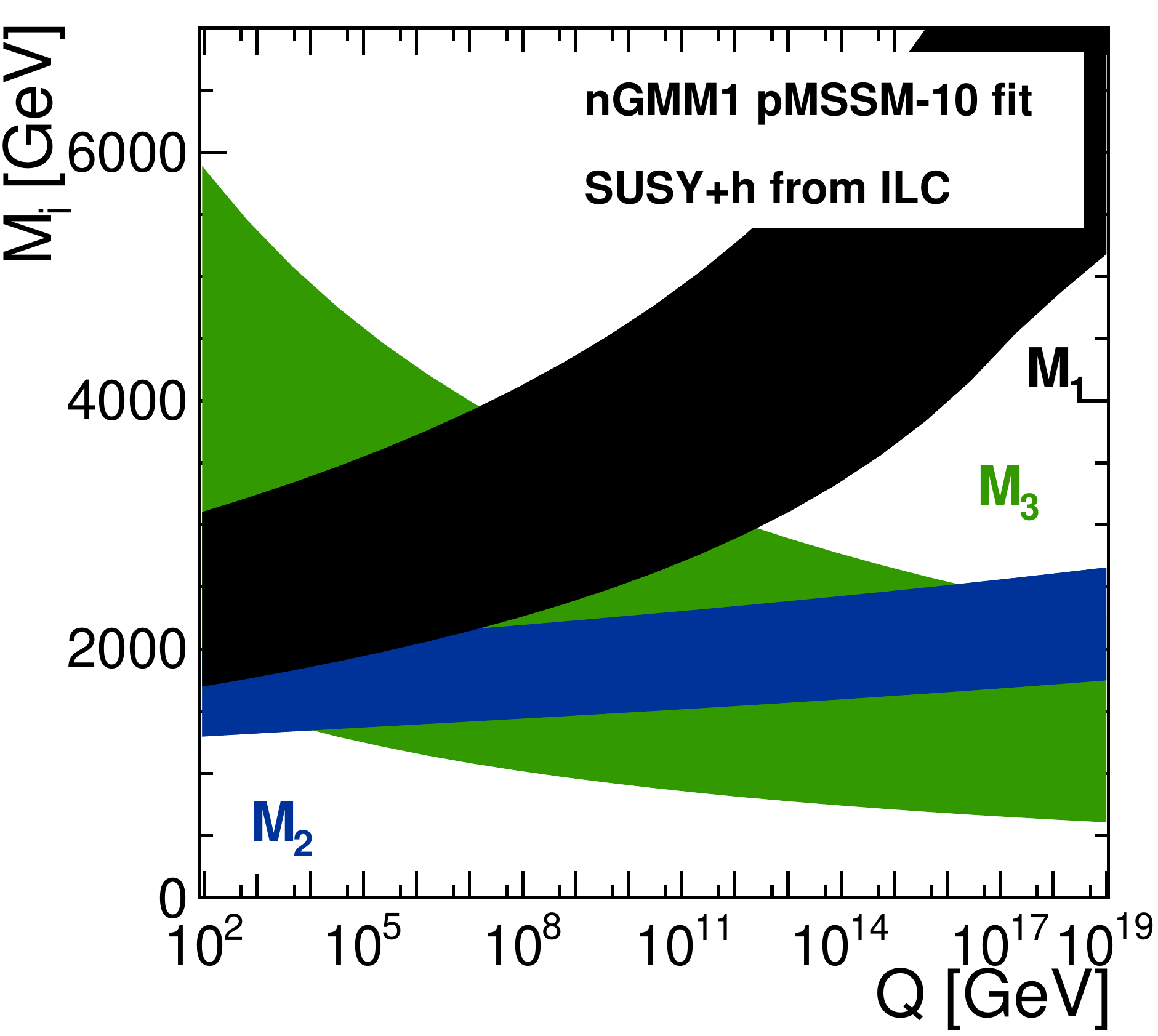}
    \caption{}
    \label{fig:RGE_nGMM1}
  \end{subfigure}
	\hfill
  \begin{subfigure}{.32\linewidth}
    \includegraphics[width=\textwidth]{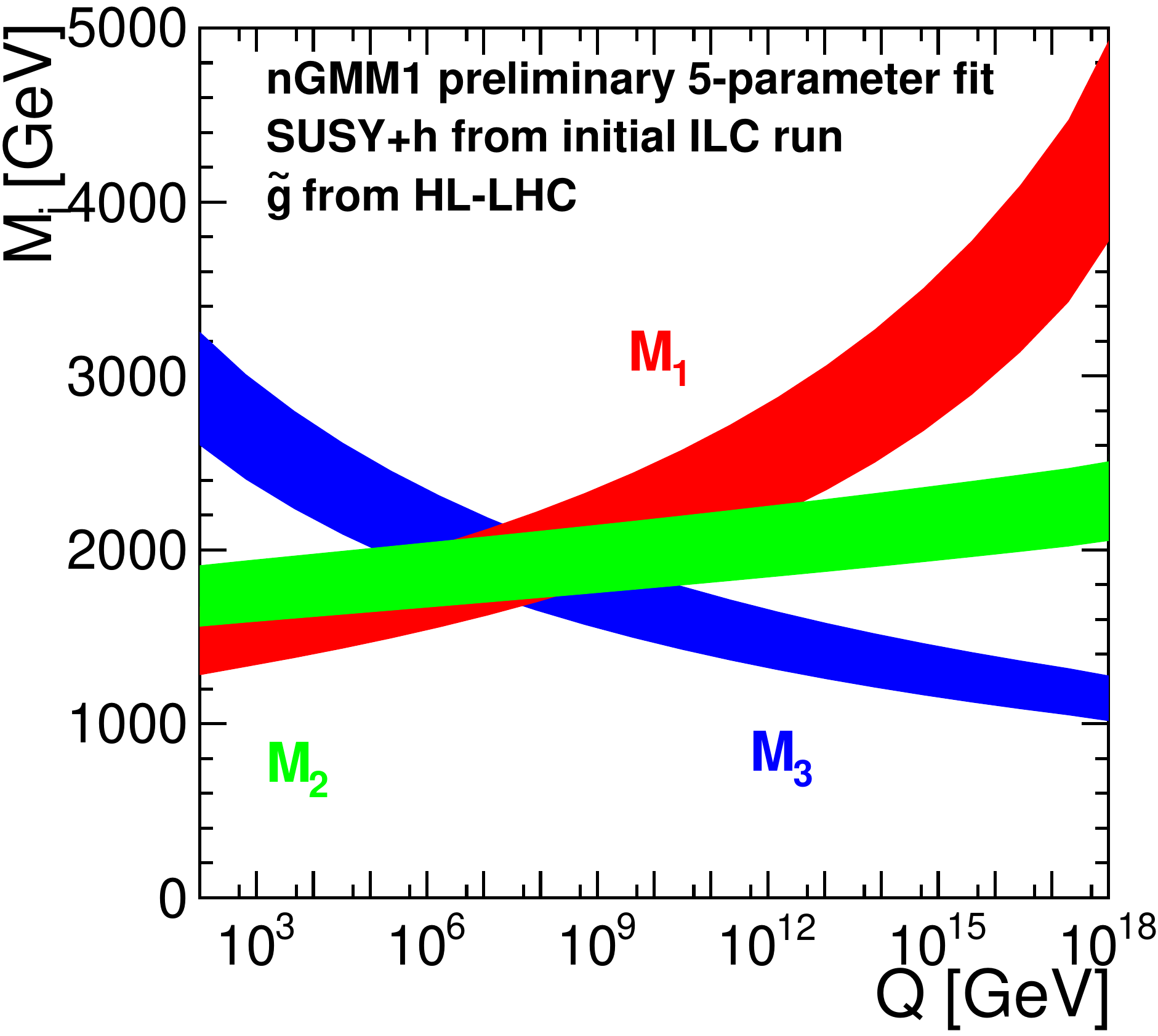}
    \caption{}
    \label{fig:RGE_nGMM1_hypot}
  \end{subfigure}
  \caption{Light higgsinos with few-GeV mass differences at the ILC: (a) RGE running of gaugino mass parameters determined from a weak-scale pMSSM-10 fit to ILC measurements in an NUHM2 benchmark. $M_3$ is extrapolated back down to the weak scale assuming gaugino mass unification, predicting a gluino at about 3\,TeV. (b) Same excercise in case of a mirage mediation benchmark. A standard GUT-scale unification of the gaugino masses can be excluded at the $99.9\%$ confidence level. (c) Same benchmark as in (b), but with more optimistic assumptions described in the text. Figures taken from Ref.~\cite{Lehtinen:2018PhD}.}
  \label{fig:gut}
\end{figure}

Figure~\ref{fig:gut} goes, however, even one step further: it shows the RGE running of the gaugino mass parameters as determined from a weak-scale pMSSM-10 fit to ILC observables up to the GUT scale. The width of the bands reflects both the uncertainties of the parameter values at the weak scale as well as the uncertainty in the RGE running due to the finite knowledge of the involved SUSY parameters. Three different cases are displayed: Fig.~\ref{fig:RGE_ILC2} shows the situation in the NUHM2 benchmark with the medium mass differences based on full simulation of the ILD detector.  $M_3$ is extrapolated back down to the weak scale assuming gaugino mass unification at the GUT-scale, with $M_{1/2}$ being determined to about 10\% precision from  $M_1$ and $M_2$. The smallest mass difference case has been studied in a benchmark based on a mirage mediation model and is shown in Fig.~\ref{fig:RGE_nGMM1}. A standard GUT-scale unification of the gaugino masses can be excluded at the $99.9\%$ confidence level. 
Figure~\ref{fig:RGE_nGMM1_hypot} shows the RGE running in the same benchmark, but instead of using the mass and cross section precision obtained from the full simulation study, hypothetical resolutions of 1\% on the masses and 3\% on the cross sections have been used. This corresponds to about a factor of 2 improvement in the experimental resolutions. In addition, a 10\% measurement of the gluino mass at the HL-LHC has been assumed here, and the parameters for the multi-TeV scalars have been fixed. As can be seen from the comparison of Figs.~\ref{fig:RGE_nGMM1} and~\ref{fig:RGE_nGMM1_hypot},
such improvements would make a qualitative difference to the determination of the mass unification scale, where the impact of the fixed parameters remains to be quantified. Possibilities to improve the capabilities of the ILD detector are being evaluated by the ILC concept group.

\subsection[Stau co-annihilation models]{$\tilde{\tau}$ co-annihilation models}
A long term focus of the SFB-B1 project has been the study of SUSY scenarios with a $\tilde{\tau}$ as the next-to-lightest SUSY particle (NLSP), especially when its mass difference to the lightest SUSY particle (LSP) is small. In these cases, the cosmologically observed relic density of dark matter can be explained by co-annihilation between NLSP and LSP.
At the same time, the searches for $\tilde{\tau}$'s are very challenging at the LHC, so that currently there is effectively no exclusions beyond the LEP results~\cite{Aad:2015eda, CMS:2017rio}, and also the HL-LHC prospects are very limited~\cite{ATL-PHYS-PUB-2016-021}, in particular for mass differences less than about 100\,GeV.

\begin{figure}[tb]
\centering
  \begin{subfigure}{.485\linewidth}
    \includegraphics[width=\textwidth]{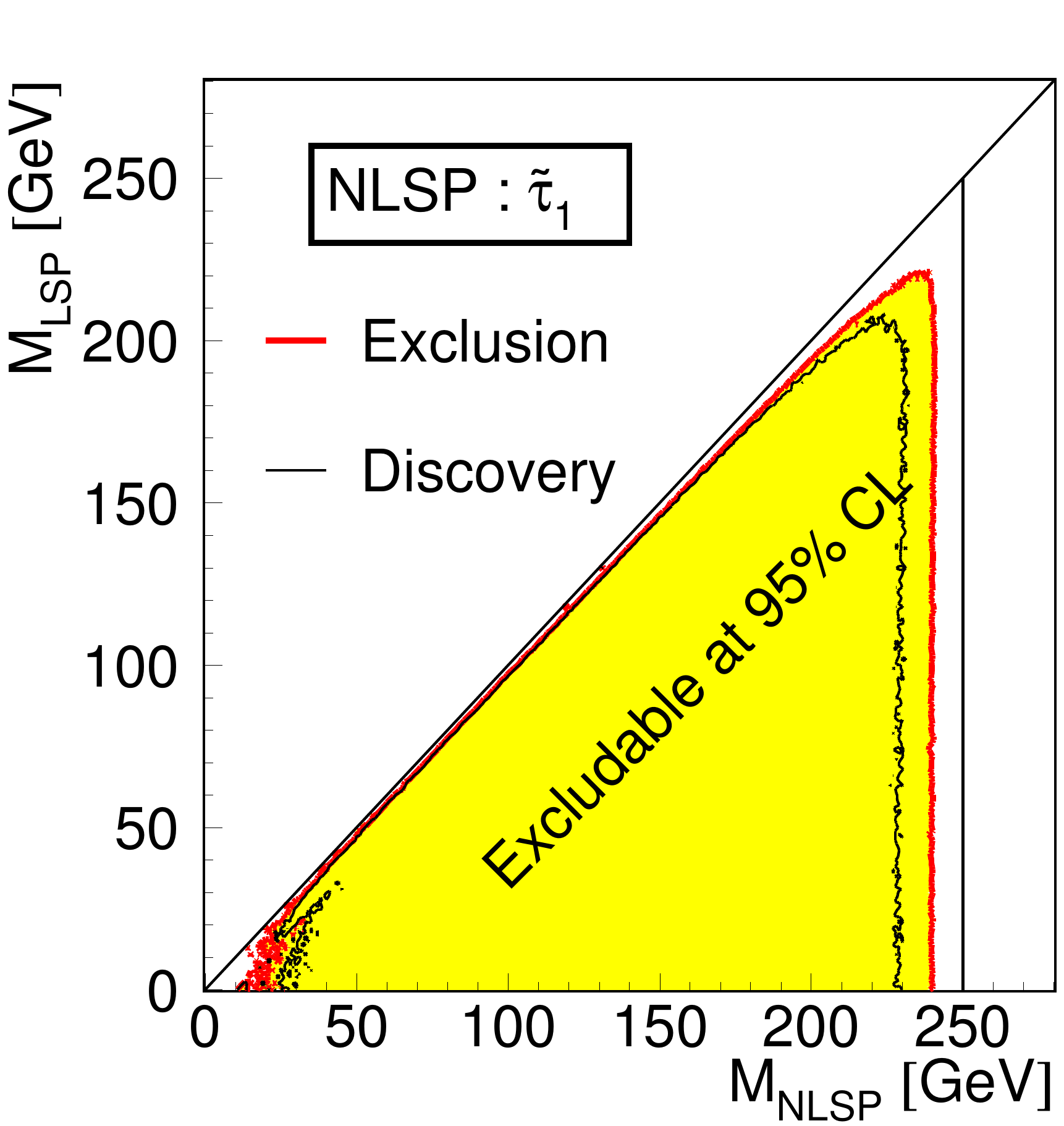}
    \caption{}
    \label{fig:staureach}
  \end{subfigure}
	\hfill
  \begin{subfigure}{.495\linewidth}
    \includegraphics[width=\textwidth]{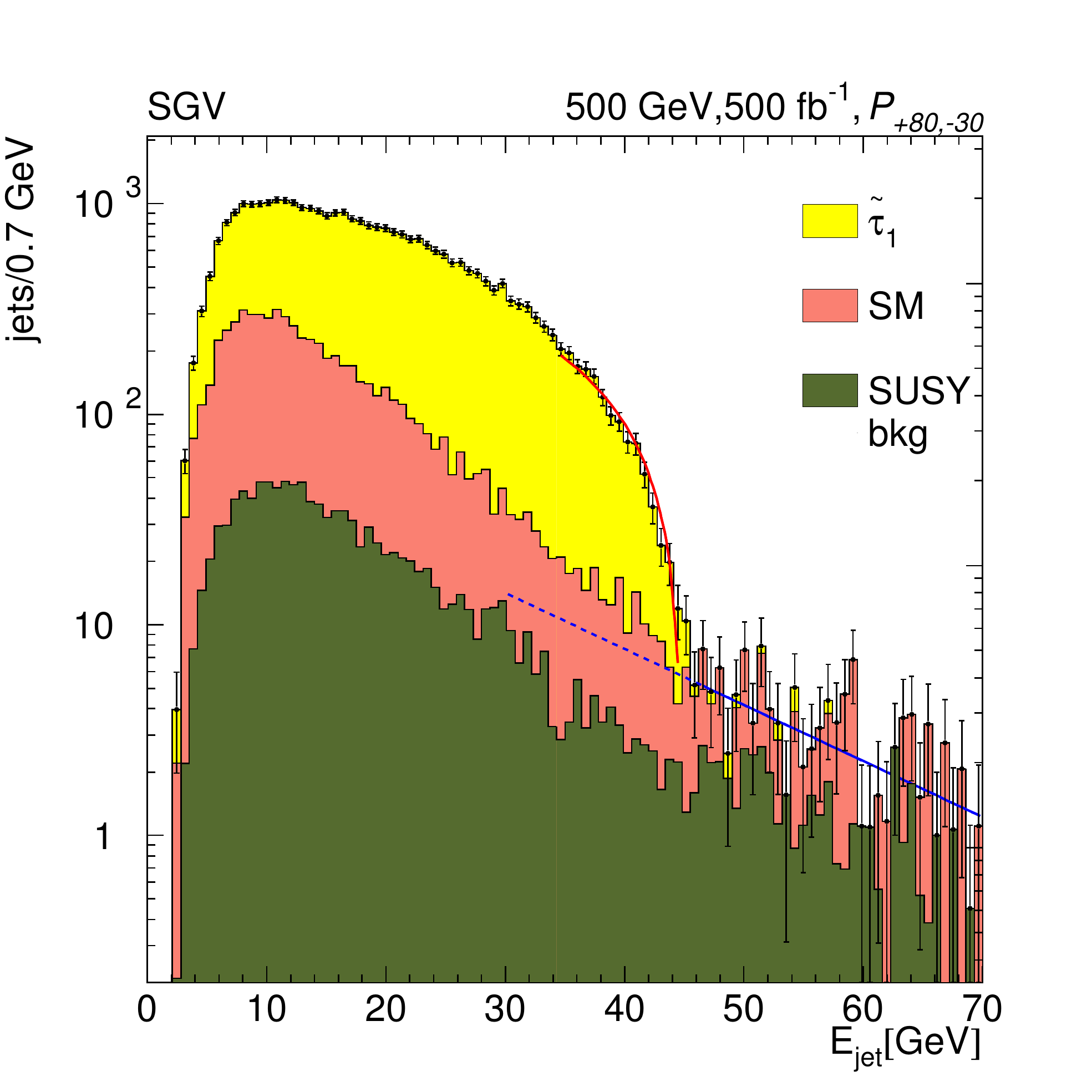}
    \caption{}
    \label{fig:stau1}
  \end{subfigure}
  \caption{ILC capabilities in the $\tilde{\tau}$ sector from a simulation study of $\tilde{\tau}_1$ pair production based on the ILD detector: (a) exclusion and discovery reach in the $\tilde{\tau}$ vs LSP mass plane for 500\,fb$^{-1}$ at 500\,GeV, corresponding to about an eighth of the foreseen data set at 500\,GeV. Figure taken from Ref.~\cite{Berggren:2013vna}. (b) $\tau$ energy spectrum from signal and SUSY and SM backgrounds. The measurement of the endpoint enables a determination of the $\tilde{\tau}$ mass with a precision of 200\,MeV. Figures taken from Ref.~\cite{Berggren:2015qua} as update of Ref.~\cite{Bechtle:2009em}.}
  \label{fig:stau}
\end{figure}

The prospects for discovering any NLSP, but especially also the most difficult case of a $\tilde{\tau}$ NLSP at the ILC have been evaluated in~\cite{Berggren:2013vna}.
As can be seen in Fig.~\ref{fig:staureach}, the exclusion and discovery potential in the $\tilde{\tau}$ vs LSP mass plane is highly complementary to (HL-)LHC prospects, since also compressed spectra can be probed up to a few GeV below the kinematic limit already with about an eighth of the total luminosity. This immediately leads to the question of precision spectroscopy, which has been studied originally based on the famous SPS1a' benchmark scenario~\cite{Bechtle:2009em, Bechtle:2009ty}. These studied showed that production cross sections and masses of $\tilde{\tau}_1$ and $\tilde{\tau}_2$ can be measured to the percent-level, and that the polarisation of the $\tau$'s from the $\tilde{\tau}_1$ decay, which gives a handle on the mixings of the $\tilde{\tau}$ and the LSP, can be determined to a few percent. Figure~\ref{fig:stau1} shows the $\tau$ energy spectrum as expected from the $\tilde{\tau}_1$ decay which is input to the determination of the $\tilde{\tau}_1$ mass via kinematic edges.

The original SPS1a' scenario had a very light coloured sector, which was quickly excluded after the start of the LHC. Therefore the more recent studied where based on the STC benchmark series from~\cite{Baer:2013ula} which has a very similar electroweak sector than SPS1a', but a much heavier coloured sector. In a joined study between the SFB-B1 project and the DESY CMS group, the interplay between LHC and ILC was investigated in these benchmarks, as opposed to the by then popular simplified models~\cite{Berggren:2015qua}. This study highlighted the complementarity and synergies between hadron and lepton colliders, even identifying cases where the knowledge of some sparticle masses from the ILC enables the targeted search for heavier states in LHC data. It also showed that in full SUSY models which can include many decays modes and long decay chain, the naive application of limits formulated in simplified models can be misleading. 

The final step then was to investigate whether the precisions of masses and cross sections would be sufficient to e.g.\ identify the LSP as main component of dark matter, to determine the underlying SUSY model and parameters as well as to predict masses of yet unobserved sparticles, which woud provide important input to upgrades of the ILC or the design of e.g.\ the next high-energy hadron collider~\cite{Lehtinen:2018PhD}. 


\begin{figure}[tb]
\centering
  \begin{subfigure}{.485\linewidth}
    \includegraphics[width=\textwidth]{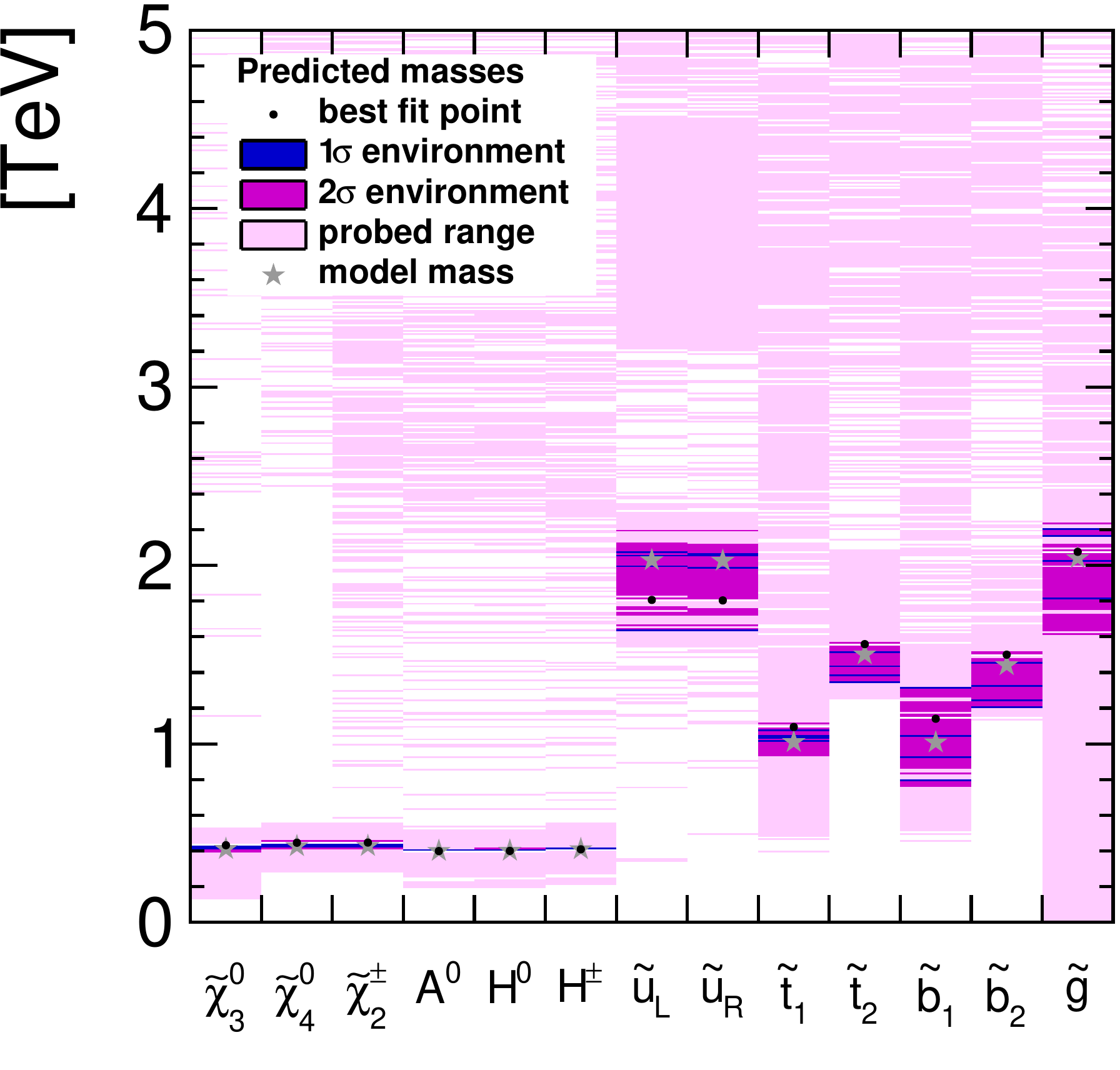}
    \caption{}
    \label{fig:STC10_massfit}
  \end{subfigure}
	\hfill
  \begin{subfigure}{.485\linewidth}
    \includegraphics[width=\textwidth]{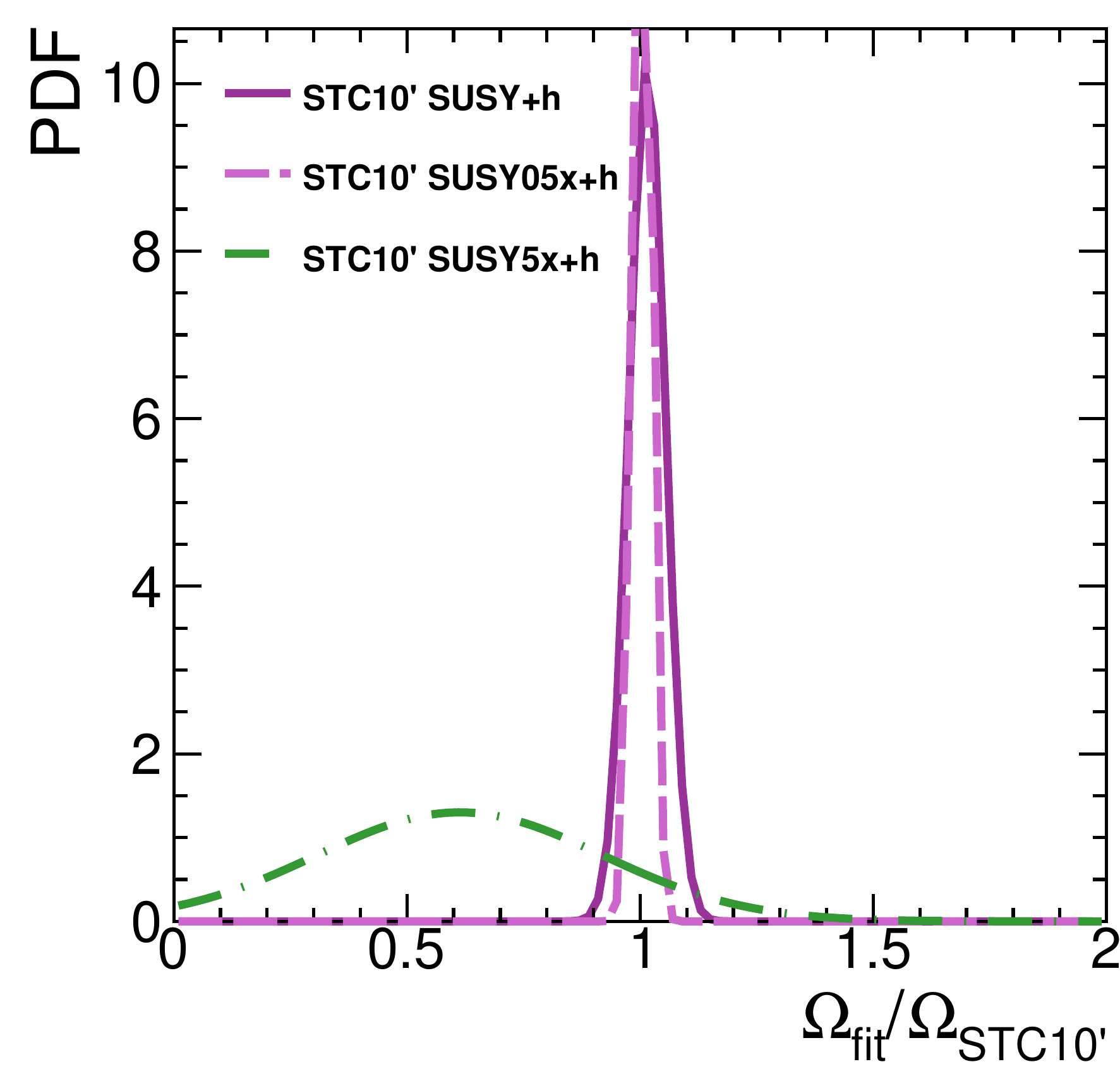}
    \caption{}
    \label{fig:STC10_omega}
  \end{subfigure}
  \caption{SUSY parameter determination at the ILC: (a) Masses of unobserved sparticles predicted from an pMSSM-13 fit to ILC SUSY and Higgs observables based on 1000\,fb$^{-1}$ at 500\,GeV in the STC10' benchmark (b) Prediction of the LSP relic density from the same fit, showing also the impact of 5 times worse and 2 times better precisions, whether the latter corresponds to the 4000\,fb$^{-1}$ at 500\,GeV in  the standard H20 running scenario. Figures taken from Ref.~\cite{Lehtinen:2018PhD}.}
  \label{fig:stcfit}
\end{figure}

Figure~\ref{fig:STC10_massfit} shows the predictions for the masses of the unobserved sparticles obtained from an pMSSM-13 fit to ILC SUSY and Higgs observables, based on a data set of 1000\,fb$^{-1}$ at 500\,GeV, split equally between ${\cal{P}}(e^-e^+)=(\pm 80\%, \mp 30\%)$, corresponding to about a quarter of the foreseen luminosity at 500\,GeV.
The probability density function for the prediction of the LSP relic density from the same fit is displayed in Fig.~\ref{fig:STC10_omega},
along-side the analogous results for a 5 times worse and a 2 times better resolution. It can be clearly seen that in the worse resolution case it is not possible to constrain the relic density in a satisfactory way, thus the precisions at the percent and for some observables at the permille are essential in order to identify the LSP as the dark matter particle.


\subsection{Impact of electroweak loops, in particular on dark matter searches}
Since the ILC is designed to perform high precision measurements of masses, cross sections and branching ratios with high accuracy 
in the  per-cent level, such a precision has to be matched from the theoretical side as well and tree-level calculations will not be sufficient. 
This project focuses on the SUSY Higgs and electroweak sector and the relevant parameters. 
Since via the loop effects not only the fundamental parameters $M_1$, $M_2$, $\mu$, $\tan\beta$ enter, but also other mass parameters of heavier states, the well-known strategies for determining the fundamental SUSY parameters~\cite{Choi:2001ww} have to be extended.
In many models, the lightest and stable SUSY particle is the neutralino $\tilde{\chi}^0_1$, often treated as suitable dark matter candidate. The precise determination of parameters has therefore direct impact on the predicted dark matter contribution as well. 

\begin{figure}[tb]
\centering
  \begin{subfigure}{.435\linewidth}
    \includegraphics[width=\textwidth]{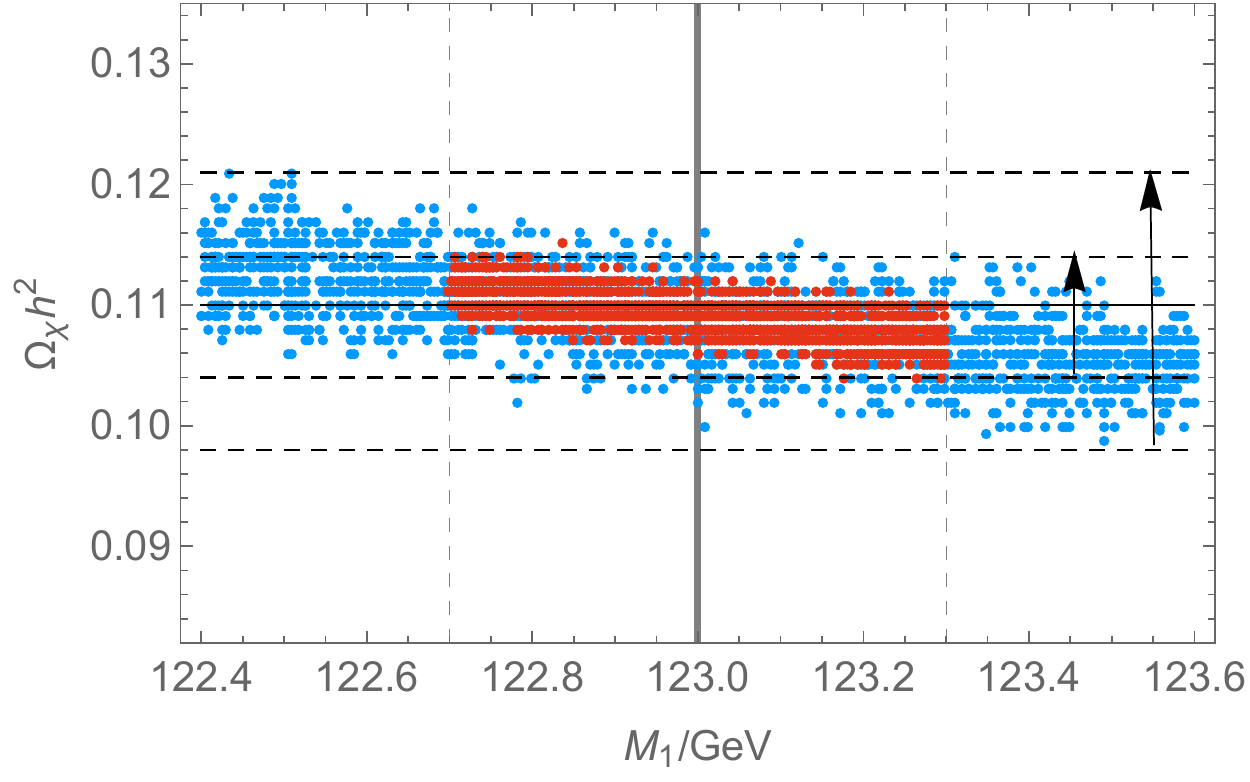}
    \caption{}
  \end{subfigure}
	\hfill
  \begin{subfigure}{.555\linewidth}
    \includegraphics[width=\textwidth]{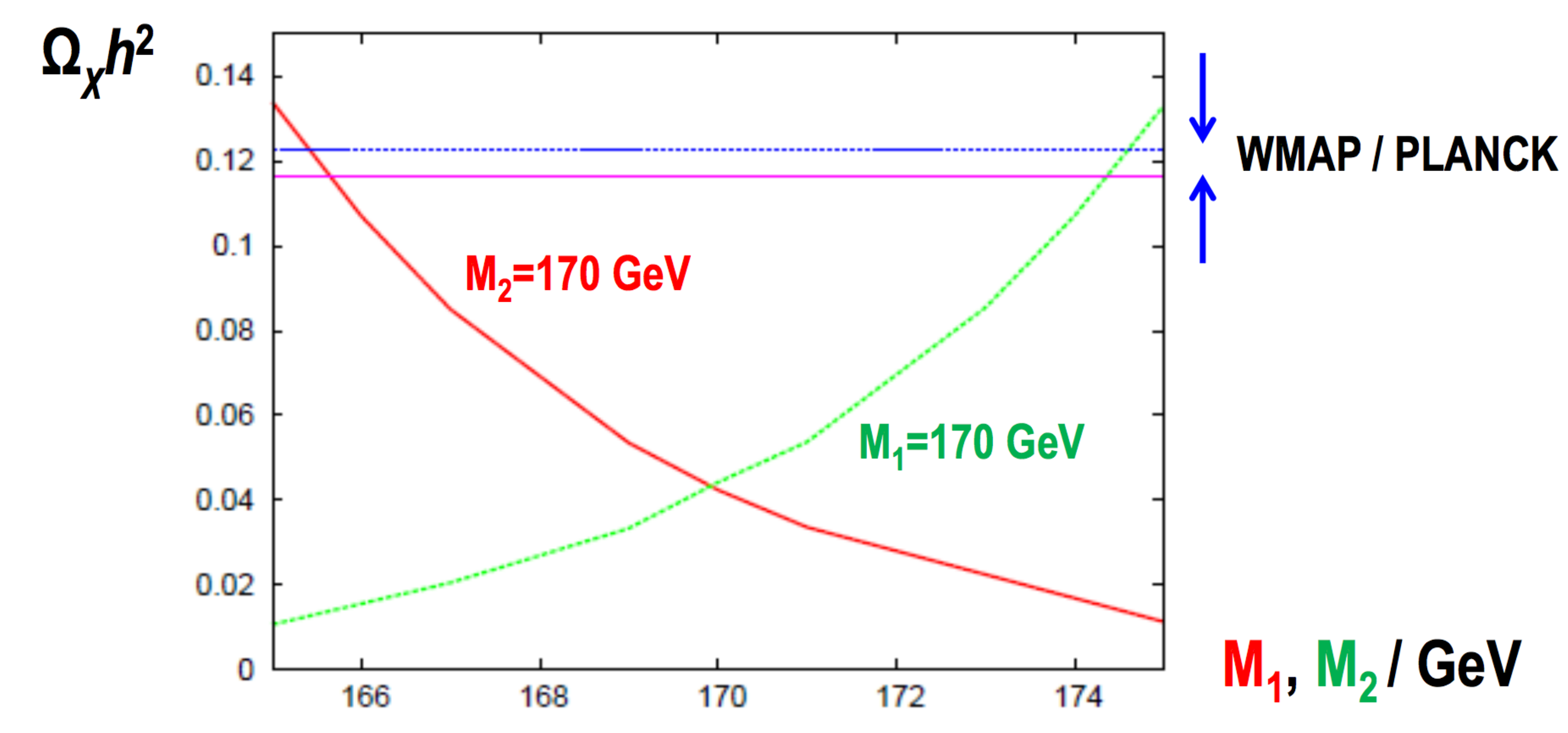}\vspace{2.4ex}
    \caption{}
  \end{subfigure}
  \caption{
(a) Determination of fundamental SUSY parameters at $\%$ level (including evaluation of loop corrections) in scenario S1
either via mass measurements in he continuum (blue area) or via threshold scans (red area) and the resulting uncertainty in the prediction of the dark matter density.
(b) Dark matter density caused by the DM candidate $\tilde{\chi}^0_1$ and its strong dependence on the gaugino 
SUSY parameters. High precision is required for accurate predictions for $\Omega h^2$~\cite{dm-eps:2015}.}
  \label{fig:aoife-dm-gmp}
\end{figure}

The first step in this project~\cite{Bharucha:2012ya,Bharucha:2012nx,Bharucha:2012wu}
 was to incorporate quantum corrections in the theoretical calculation in order to determine the underlying SUSY parameters from measurements of chargino/neutralino masses and cross sections. The one-loop predictions were fitted to the prospective measurements of cross sections, forward-backward asymmetries and the accessible chargino and neutralino masses. Since the 
 one-loop contributions are dominated by the stop sector, the accurate determination of the desired parameters provides also access to the stop masses and mixing angle. Having determined the fundamental 
 parameters~\cite{Bharucha:2012qr,Bharucha:2012pg,Bharucha:2013jha} using 1-loop corrected observables, we have applied the results to the prediction of the expected dark matter contribution. The impact of the one-loop corrected parameters, in particular $M_1$ causes up to a 10\% correction for the corresponding dark matter contribution in $\Omega h^2$, see 
 Fig.~\ref{fig:aoife-dm-gmp}. 
 
A particular challenge of this SFB project \cite{Bharucha:2012nx} was to work out consistent renormalization in the on-shall scheme in the complex MSSM. The impact of this quantum corrections has been evaluated  in Higgs decays $h(a) \to \tilde{\chi}_i^+\tilde{\chi}_j^-$, where the Higgs-propagator corrections have been incorporated up to the two-loop level, as well as in 
chargino production $e^+ e^- \to \tilde{\chi}_i^+ \tilde{\chi}_j^-$. Concerning the parameter renormalisation in the chargino and neutralino
sector, we have shown that the phases of the parameters in the chargino and
neutralino sector do not need to be renormalised at the one-loop level. We
have therefore adopted a renormalisation scheme where only the absolute
values of the parameters $M_1$, $M_2$ and $\mu$ are subject to the renormalisation
procedure. In order to perform an on-shell renormalisation for those parameters
we have worked out the strategy
choosing three out of the six masses in the chargino and
neutralino sector that are renormalised on-shell, while the predictions for the
physical masses of the other three particles receive loop corrections. 

\subsection{Extended SUSY models}

SFB676-B1 project studied also extended SUSY models, as for instance, unification models, models with an extended $U(1)$ sector as well as models with R-parity violation that can explain neutrino mixing scenarios.

\subsubsection{Unification models: supersymmetric SO(10) models}
\begin{figure}[tb]
\centering
  \begin{subfigure}{.55\linewidth}
    \includegraphics[width=\textwidth]{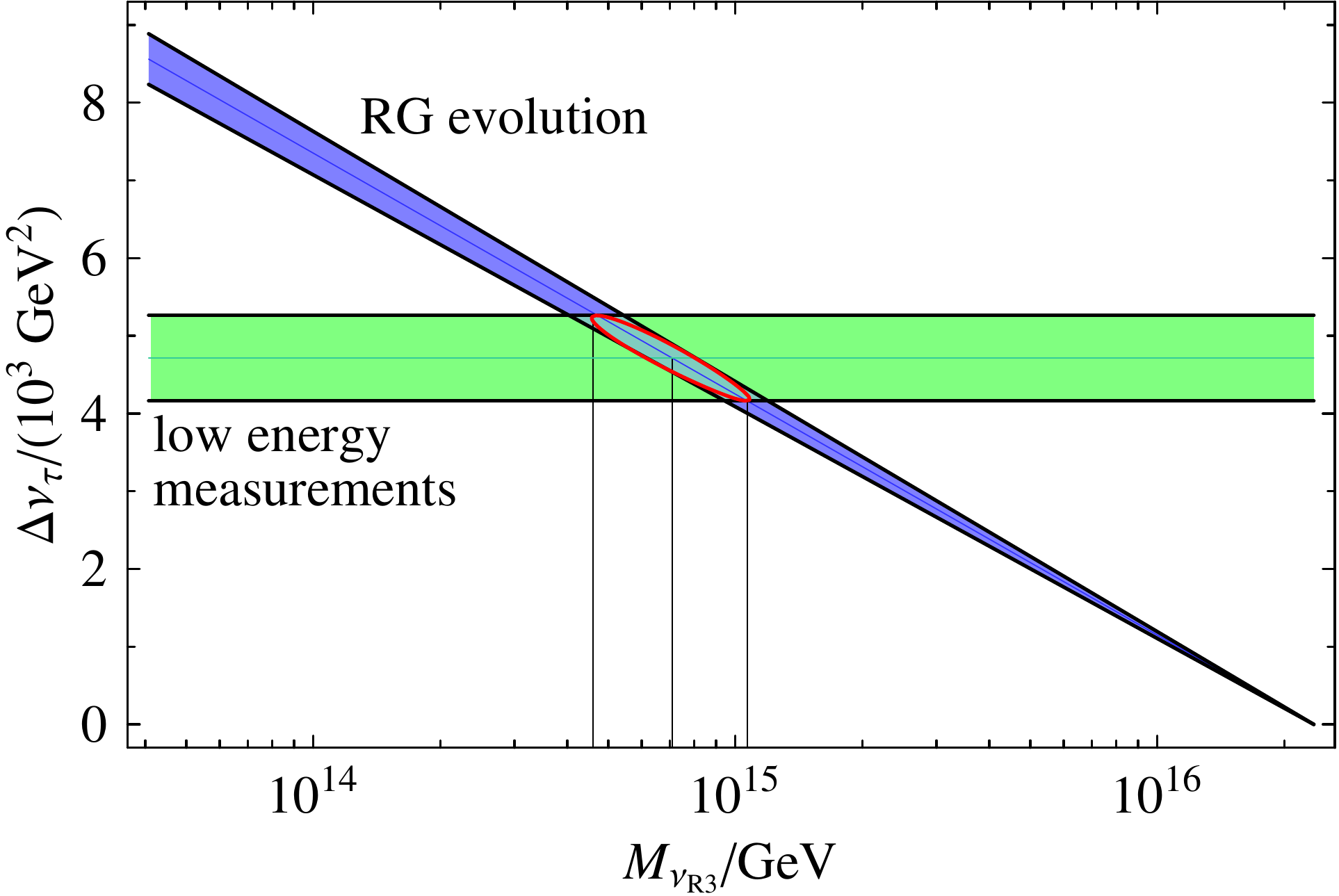}
    \caption{}
  \label{fig:deppisch-gmp}
  \end{subfigure}
	\hfill
  \begin{subfigure}{.44\linewidth}
    \includegraphics[width=\textwidth]{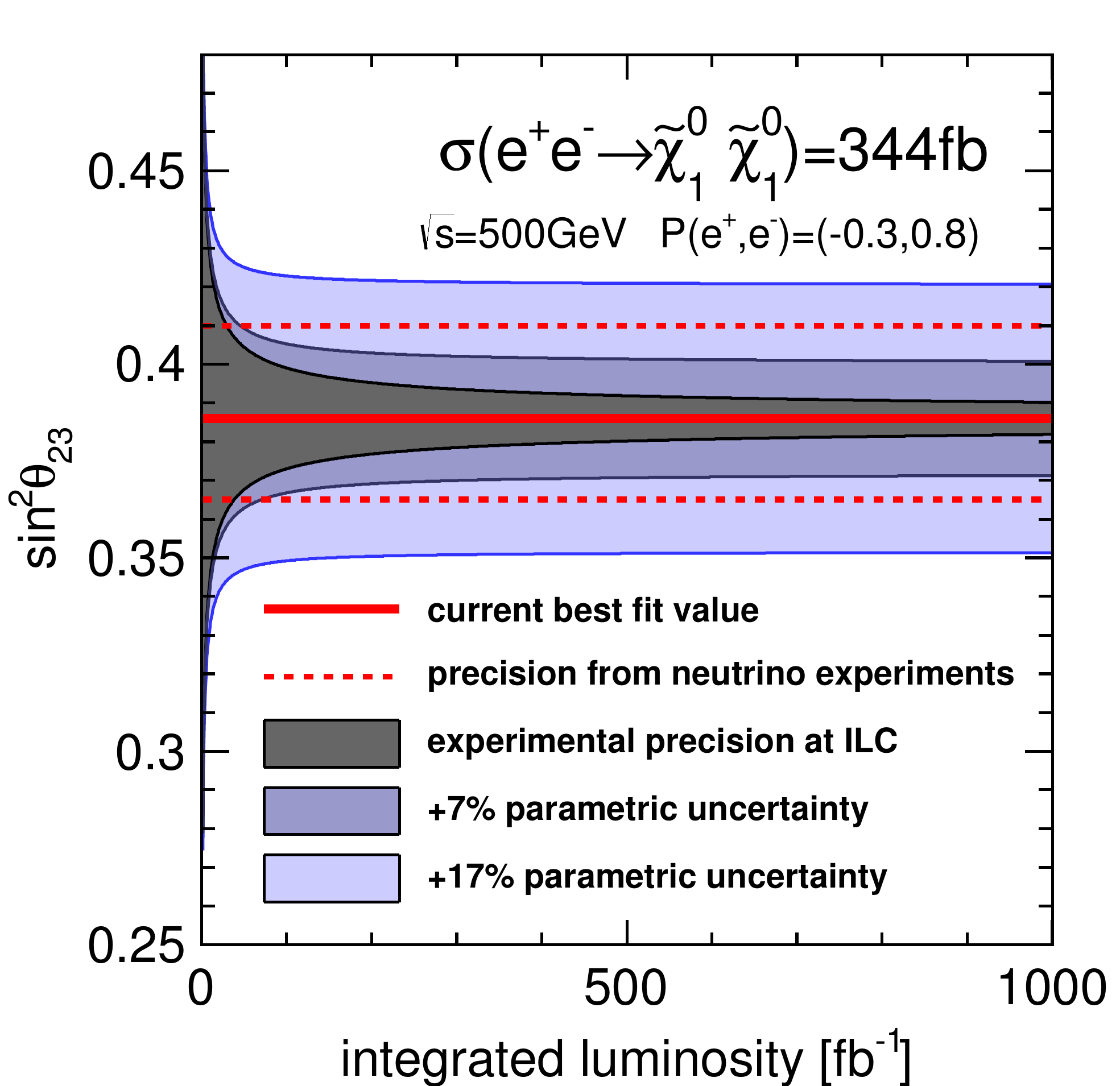}
    \caption{}
  \label{fig:bRPV}
  \end{subfigure}
  \caption{(a) 
Shift $\Delta_{\nu\tau}$ of the third generation L slepton mass parameter generated by loops involving heavy neutrino R-superfields. The blue area corresponds to the prediction originating from
the renormalization group [RG], whereas the green band is determined by low-energy mass
measurements. 
Reprinted figure with permission from Ref.~\cite{Deppisch:2007xu}. Copyright (2008) by the American Physical Society.
(b) Precision of the measurement of the neutrino mixing angle $\theta_{23}$ via 
$e^+e^-\to \tilde{\chi}^0_1 \tilde{\chi}^0_1 \gamma$ at the ILC compared with the corresponding achievable precision via 
neutrino experiments. Figure taken from Ref.~\cite{List:2013dga}.
}
\end{figure}

Extrapolations of soft scalar mass parameters in supersymmetric theories can be used to explore elements of the physics scenario near the grand unification scale. In \cite{Deppisch:2007xu} the potential of this method in the lepton sector of SO(10) which incorporates right-handed neutrino superfields has been explored. Two examples have been analyzed in which high-scale parameters in supersymmetric SO(10) models have been connected with experimental observations that could be
expected in future high-precision Terascale experiments at LHC and $e^+e^-$ linear colliders: a)
the case for a one-step breaking SO(10) $\to$ SM and b)  the analysis of two-step
breaking SO(10) $\to$ SU(5) $\to$ SM. The renormalization group provides the tool for bridging the gap between
the Terascale experiments and the underlying high-scale grand unification theory. Even
though it depends on the detailed values of the parameters with which resolution the highscale
picture can be reconstructed, a rather accurate result could be established in the example
for one-step breaking SO(10) $\to$ SM, including the heavy mass of the right-handed neutrino
$\nu_{R3}$ expected in the seesaw mechanism. As naturally anticipated, the analysis of two-step
breaking SO(10) $\to$ SU(5) $\to$ SM turns out to be significantly more difficult, demanding a
larger set of additional assumptions before the parametric analysis can be performed, see Fig.\ref{fig:deppisch-gmp}. 

\subsubsection{U(1)-extended SUSY models: light singlets}
Motivated by grand unified theories and string theories, there has been analyzed in \cite{Choi:2006fz} 
the general structure of the neutralino sector in the USSM, an extension of the Minimal Supersymmetric Standard Model that involves a broken extra U(1) gauge symmetry.
This supersymmetric U(1)-extended model includes an Abelian gauge superfield and a Higgs singlet superfield in addition to the standard gauge and Higgs superfields of the MSSM. The interactions between the MSSM fields and the new fields are in general weak and the mixing is small, so that the coupling of the two subsystems can be treated perturbatively. Light singlets  escaping LEP and LHC bounds, are natural in these models. Prospects for the production channels in cascade decays at the LHC and pair production at $e^+ e^-$ colliders have been discussed in this SFB study.

\subsubsection{R-Parity violating SUSY: bridge from collider to neutrino physics}
Supersymmetry (SUSY) with bilinearly broken R parity (bRPV)  does not provide any dark matter candidates, but offers an attractive possibility to explain the origin of neutrino masses and mixings. In such scenarios, the study of neutralino decays at colliders can 
give access to neutrino parameters. In \cite{List:2013dga} a full detector simulation for 
neutralino $\tilde{\chi}^0_1$-pair-production and two-body decays has been performed including Standard Model background processes. As studied parameter point a worst case scenario has been used, where
$m_{\tilde{\chi}^0_1}\sim m_{W/Z}$, thus, the signal significantly overlaps with SM background. It has been developed a model-independent selection strategy to disentangle the different event classes involving the two decay modes of the LSP 
$\tilde{\chi}^0_1 \to \mu^{\pm} W^{\mp}$  and $\tilde{\chi}^0_1 \to \tau^{\pm} W^{\mp}$.
The $\mu\mu$ and $\mu\tau$ events have been used to determine the ratio of the two branching
ratios $BR(\tilde{\chi}^0_1\to \mu^{\pm}W_{\mp})/ 
BR(\tilde{\chi}^0_1\to \tau^{\pm}W_{\mp})$, which is related to the atmospheric neutrino
mixing angle $\sin^2\theta_{23}$. For an integrated luminosity of 500~fb$^{-1}$ the total uncertainty on
this ratio, including statistical and systematic uncertainties, has been determined to 4\%, see Fig.\ref{fig:bRPV}.
 In addition it has been shown that the precision in measuring the atmospheric neutrino mixing
angle is in the same range than measurements from neutrino oscillation experiments, even
when taking parametric uncertainties due to the unknown parts of the SUSY spectrum
into account. Therefore, the International Linear Collider is highly capable to test bRPV
SUSY as origin of neutrino masses and mixings.

\section{Other extensions of the SM}
Although Supersymmetry is one of the best motivated BSM models, explaining several of the remaining open questions of the SM, it is of great importance to study also further BSM models like Little Higgs models  as well as models with large extra dimensions 
 or gauge group extended models.



\subsection{Automatisation of BSM models}
It is of great importance to check  new physics models against data of the LHC and of expected data at the ILC.
The original purpose of the computing tool CheckMATE, public code to perform collider phenomenology, was to offer theorists a way
to quickly test their favourite BSM models against various existing LHC analyses. It consists of an automatised chain of Monte Carlo event generation, detector simulation, event analysis and statistical evaluation and allows to check whether a given parameter point of a BSM model is excluded or not on basis of currently more than 50 individual ATLAS or CMS analyses at both energies 
$\sqrt{s}=8$ and 13 TeV~\cite{Dercks:2018xyk}. 
The used recasting procedure is a powerful procedure which applies existing
collider results on new theoretical ideas without
requiring the full experimental data analysis to be
restarted. Theories which share experimentally indistinguishable topologies can be tested via identical event
selection techniques, bearing the advantage that background
expectation and the number of observed events
stay constant.

We were able to make use of the
powerful features of the Monte Carlo event generator
Whizard to simulate $e^+ e^-$ collisions including
the effects of beam polarisation, initial state radiation
and beamstrahlung. Furthermore, we continue using the
fast detector simulation Delphes to describe the ILD detector.
Currently it provides a good approximation
of the most relevant acceptance and efficiency factors.
Lastly, we extended the set of accessible parameters in
CheckMATE, allowing users to test different polarisation
and luminosity combinations. This makes it very convenient
to discuss the importance of e.g.\ the lepton polarisation
for the overall sensitivity of the experiment to a
given BSM hypothesis.
CheckMATE fuctionality has been extended to analysis studies for different ILC set-ups, as, for instance, different beam polarization configurations, variable energies 
etc. Of particular interest are challenging studies for the LHC, as, for instance, monophoton searches for dark matter particles in compressed spectra~\cite{Dercks:2018hgz}. 

\subsection{Little Higgs models}

Another model that has been checked with the help of CheckMATE
against real LHC data at 8~TeV and13 TeV  are little Higgs models with T-Parity \cite{Dercks:2018hgz}.
We scrutinize the allowed parameter space of Little Higgs models with the concrete symmetry of T-parity by providing comprehensive analyses of all relevant production channels of heavy vectors, top partners, heavy quarks and heavy leptons and all phenomenologically relevant decay channels. 

This model is an elegant implementation of global collective symmetry breaking combined
with a discrete symmetry to explain the natural lightness of the Higgs boson as a (pseudo-) Nambu-Goldstone boson.
This model predicts heavy partners for the Standard
Model quarks $q_H$, leptons $\ell_H$, gauge bosons $W_H$, $Z_H$, $A_H$ and special partners for the top quark $T^\pm$.
Constraints on the model will be derived from the signatures of jets and missing energy or leptons and missing energy. Besides the symmetric case, we also study the case of T-parity violation. Furthermore, we give an extrapolation to the LHC high-luminosity phase at 14 TeV. 
Our results show that, although the Littlest Higgs model with T-parity has been constrained much stronger by
LHC run 2 data, it is still a rather natural solution to the shortcomings of the electroweak
and scalar sector, and we will need full high-luminosity data from the LHC to decide whether
naturalness is actually an issue of the electroweak sector or not. A qualitative improvement
of all bounds on the model, particularly in the Higgs sector and the heavy lepton sector,
might need the running of a high-energy lepton collider (or a hadron collider at much higher
energy).


\subsection{Z' models, models with large extra dimensions and contact interaction models}
Another class of new physics models are models with an additional $Z'$ boson. Unfortunately up to now no additional $Z-$boson has been found (yet) at LHC. That means, the limits on $m_{Z'}$ are already in the multi-TeV region, depending on the model and on the number of extra dimensions, and it is crucial to determine the sensitivity to such new physics models  via 
measuring
 deviations of the cross sections from their Standard Model predictions.
Due to the clean environment at the ILC, precise measurements of well-known SM processes, as, for instance $W^+W^-$-production and the $Z$-pole,  are predestinated for such searches.

\subsubsection{Z' models and models with heavy leptons}
\begin{figure}[tb]
\centering
  \begin{subfigure}{.475\linewidth}
    \includegraphics[width=\textwidth]{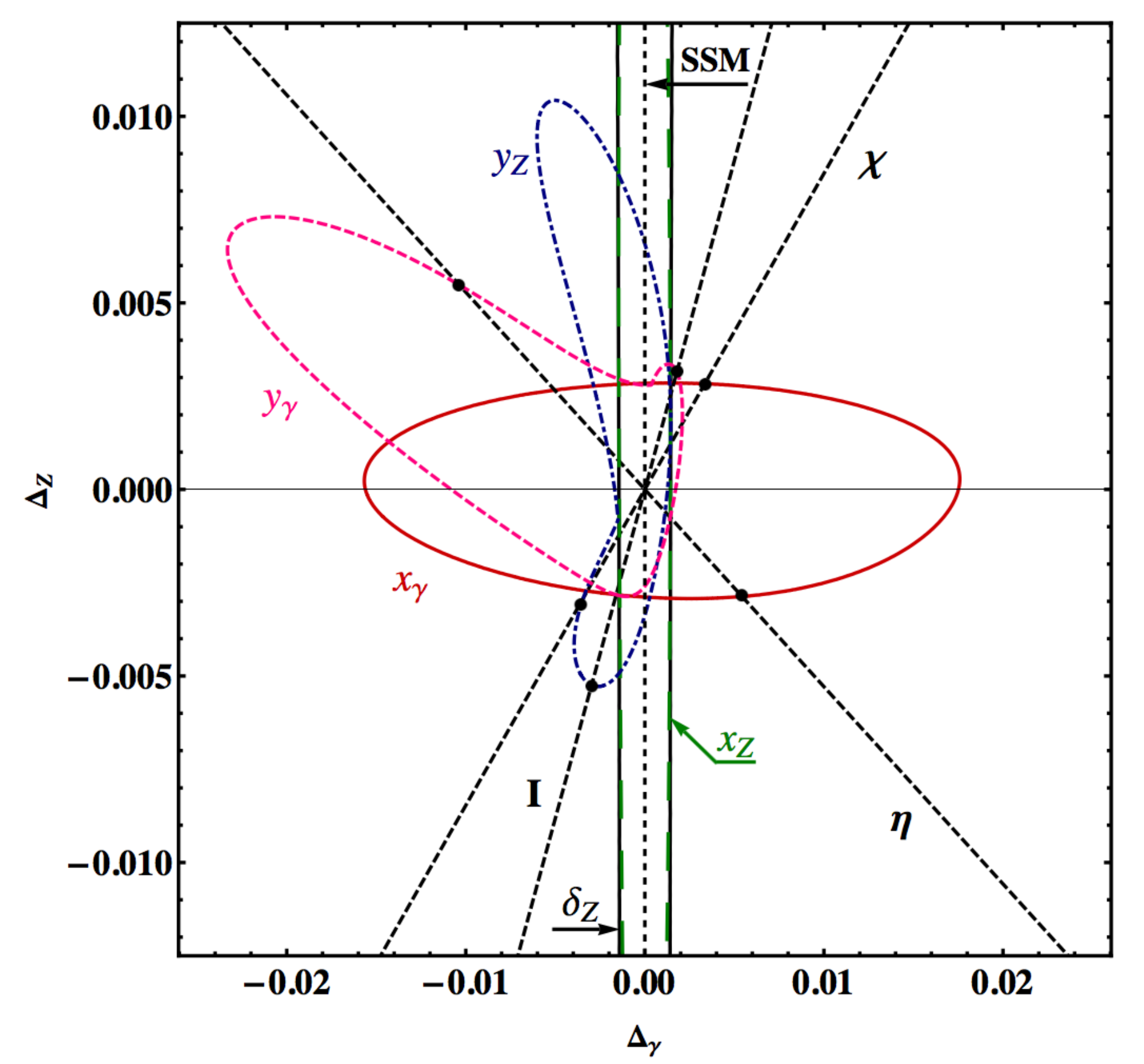}
    \caption{}
  \end{subfigure}
	\hfill
  \begin{subfigure}{.495\linewidth}
    \includegraphics[width=\textwidth]{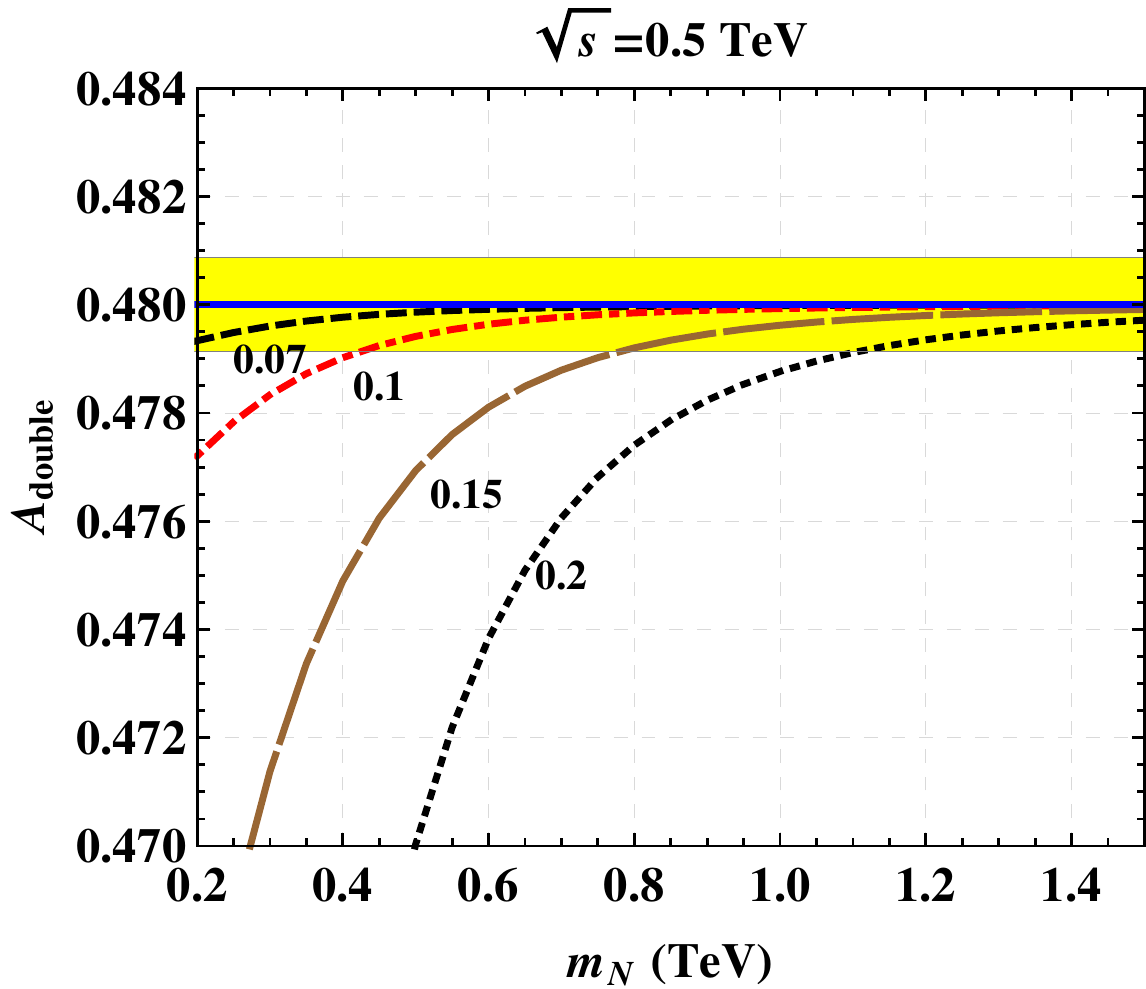}
    \caption{}
  \end{subfigure}
  \caption{(a) Generic Z' models and anomalous gauge coupling models: correponding parameters are assumed to take non-vanishing values, one at a time only: $x_{\gamma},$ $x_Z$, $y_{\gamma}$, $y_Z$ and $\delta_Z$. Dashed straight lines correspond 
to specific Z' models ($\chi,\psi,\eta,I$ and LRS). The study was done for $\sqrt{s}=500$~GeV, ${\cal L}_{int}=500$~fb$^{-1}$
and polarized beams with $P(e^-)=\pm 80\%$, $P(e^+)=\mp 50\%$. Figure reprinted from Ref.~\cite{Andreev:2012cj} with kind permission of The European Physical Journal (EPJ).
(b)  Double beam polarization asymmety $A_{\rm double}$ in the process $e^+e^-\to W^+W^-$  as function of the neutral heavy lepton mass $m_N$ for different sizes of the couplings at $\sqrt{s}=500$~GeV and ${\cal L}_{\rm int}=1$~ab$^{-1}$. 
The horizontal line corresponds to $A_{\rm double}^{\rm SM}=A_{\rm double}^{\rm Z'}=A_{\rm double}^{\rm AGC}$ (error band for the SM case at the $1$-$\sigma$ level. 
Reprinted figure with permission from Ref.~\cite{Moortgat-Pick:2013jra}. Copyright (2013) by the American Physical Society.}
  \label{fig:pankov-gmp}
\end{figure}

In this SFB-B1 study~\cite{Andreev:2012cj}, we therefore discuss the expected sensitivity to Z's in $W^\pm$-pair production cross sections at the ILC. In  particular it is focussed on the potential for distinguishing observable effects of the Z' from analogous ones in competitor models with anomalous trilinear gauge couplings (AGC) that can lead to the same or similar new physics experimental signatures at the ILC. 
The sensitivity of the ILC for probing the Z-Z' mixing and its capability to distinguish these two new physics scenarios is substantially enhanced when the polarizations of the initial beams and the produced $W^\pm$ bosons are considered. A model independent analysis of the Z' effects in the process $e^+e^- \to W^+W^-$ allows to differentiate the full class of vector Z' models from those with anomalous trilinear gauge couplings, with one notable exception: the sequential SM (SSM)-like models can in this process not be distinguished from anomalous gauge couplings. Results of model dependent analysis of a specific Z' are expressed in terms of discovery and identification reaches on the Z-Z' mixing angle and the Z' mass, cf. Fig.~\ref{fig:pankov-gmp} (a). 

In~\cite{Moortgat-Pick:2013jra}, we extended our $Z'$ searches and explore the effects of neutrino and electron mixing with exotic heavy leptons in the process 
$e^+e^-\to W^+W^-$ within $E_6$ models, which also incorporate an additional $Z'$. 
We examine the possibility of uniquely distinguishing and
identifying such effects of heavy neutral lepton exchange from $Z-Z^0$ mixing within the same class
of models and also from analogous ones in competitor models with anomalous trilinear gauge
couplings (AGC) that can lead to very similar experimental signatures at the ILC with $\sqrt{s}=350$, 500 GeV and 1 TeV.
 A clear identification of the model
with respect to  is possible by using a certain double polarization asymmetry that requires simultaneously polarized $e^-$ and $e^+$ beams. In addition, the sensitivity of the ILC for probing exotic-lepton admixture is substantially enhanced when the polarization of the produced $W^{\pm}$ bosons is considered, cf. Fig.~\ref{fig:pankov-gmp}(b).

\subsubsection{Large extra dimension and contact interactions}
Concerning high precision measurements at the Z-pole, both high luminosity and 
the polarization of both beams are mandatory to achieve a precision in the determination of the electroweak 
mixing angle, of about one order of magnitude better than at LEP and SLC. 
In~\cite{MoortgatPick:2010zz} the  physics potential at a Z-factory (corresponding to the GigaZ-option at the ILC) has been summarized. This article explains
the fundamentals in (beam) polarization and provides an overview of the impact of these spin effects in
electroweak precision physics. Measuring the left-right-asymmetry at the Z-pole under these conditions
allows to resolve the discrepancy between the experimentally measured values of $sin^2 \theta_{\rm eff}$ derived from $A_{\rm LR}$  and from $A_{\rm FB}$. The measured value has immediate impact on predictions in the Higgs
and beyond Standard Model physics sector.

In~\cite{MoortgatPick:2010zza} the potential of polarized beams has been summarized in searches for contact interactions, for large extra dimensions and SUSY. In many case, the availability of both beams polarized in particular allows in particular the distinction of different models, as for instance, between the RS-model and the ADD-model concerning extra dimension or the model-independent
determination of specific contact interactions. 

\section[Testing of QED processes at e+ e- colliders]{Testing of QED processes at $e^+e^-$ colliders}
There is also a great potential at future high-energy $e^+e^-$ linear colliders to test nonlinear QED processes. 
Since the $e^-$ an $e^+$ beams are both of high energy as well as high intensity, the interaction zone between both beams can be used for testing  QED processes: the oncoming beam generates a strong external background field for the incoming $e^-$ beam and it has to be checked under which experimental conditions the electrons have to be treated as dressed stated, the so called Volkov states within the quantum mechanical 'Furry picture', details about this description are given, for instance, 
in~\cite{MoortgatPick:2009zz}.

Within the SFB676-B1project, the impact of such  intense beams (high luminosity, high energy) have been
analyzed in the beam-beam interaction region, where strong electromagnetic fields occur~\cite{Porto:2013pia}.
The unstable vacuum present at the interaction zone might lead to a regime of nonlinear
Quantum Electrodynamics, affecting the processes in the IP area. Such conditions therefore motivate
to calculate all probabilities of the physics processes under fully consideration of the external
electromagnetic fields affecting the vacuum.
At previous lepton colliders, the much weaker external electromagnetic fields at the IPs did not
needed to be considered apart for background processes: the first order background processes as
beamstrahlung and coherent pair production, the second order incoherent pair production as well.
At future linear colliders the external fields would be orders of magnitude higher so an estimate
of the effects on all the processes is requested. As we have shown, indeed, the $\chi$ parameter, that
encodes the dependence of the probabilities on the intensity of the external field at the IP, is up to
3 orders of magnitude higher at ILC and CLIC than at LEP. In particular at CLIC-3TeV, we would
have $\chi_{\rm av} \sim 3:34$, describing a critical regime, see Tab.\ref{table:porto-gmp}.

In~\cite{Hartin:2013dca} we compared the Furry picture (FP), which separates the external field from the interaction Lagrangian and requires solutions
of the minimally coupled equations of motion, with the Quasi-classical Operator method (QOM), which is an alternative
theory in the Heisenberg picture which relies on the limiting case of ultra-relativistic particles.
We applied these theories to collider phenomenology for future linear colliders in which
two strong non-collinear fields (those of the colliding charge bunches) are present at the interaction
point. The FP requires new solutions of the equations of motion, external field propagators and
radiative corrections, in the two non-collinear electromagnetic fields. The FP applies to all physics
processes taking place at the interaction point. The QOM, however, we showed can be applied to a subset of processes
in which the quasi-classical approximation is valid~\cite{Mueller:master}. 

In~\cite{Akal:2016vtq}, we presented a comprehensive investigation of nonlinear lepton-photon interactions in
external background fields. The considered strong-field processes were Compton scattering
and stimulated electron-positron pair production (nonlinear Breit-Wheeler process).
We discuss nonlinear
Compton scattering in head-on lepton-photon collisions extended properly to beyond the
soft-photon regime.
A semi-classical method based on coherent states of radiation allowed us to treat the external
background quasi-classically in the ordinary QED action. We discussed in great detail the relevance of these extra terms by applying our
general formula to Compton scattering by an electron propagating in a laser-like background
and compared our unconstrained phase-space integrand with the one in the soft photon limit. We
showed that already the leading term in the soft limit is not sufficient to describe the exact
total scattering probability for large energies. 


\begin{table}
\centering
\begin{tabular}{|c||c|c|c|c|}
\hline  
Machine & LEP II & SLC & ILC-1TeV& CLIC-3TeV\\ \hline\hline
 Energy (GeV) & 94.5 & 46.6 & 500 & 1500\\ \hline
 $N$ (10$^{10}$) & 334 & 4 & 2 & 0.37 \\ \hline
 $\sigma_x,\sigma_y$ ($\mu$m) & 190, 3 & 2.1, 0.9 & 0.335, 0.0027 & 0.045, 0.001\\ \hline
 $\sigma_z$ (mm) & 20 & 1.1 & 0.225& 0.044\\ \hline\
 $\chi_{average}$  & 0.00015 & 0.001 & 0.27 & 3.34\\ \hline
 $\chi_{max}$  & 0.00034 & 0.0019 & 0.94 & 10.9\\ \hline
\end{tabular}
\caption{Lepton colliders parameters. $N$ is the number of leptons per bunch, $\sigma_x$, $\sigma_y$ are the transversal
dimensions of the bunches, $\sigma_z$ presents the longitudinal dimension. $E$ is the energy of the particles in the
bunches. The parameters for ILC-1TeV are taken from a 2011 dataset. Table taken from Ref.~\cite{Porto:2013pia}.
}
\label{table:porto-gmp}
\end{table}

%

\section{Conclusion} 

In conclusion, this SFB made over the 12 years of its lifetime a huge impact to the evaluation of the physics potential of future electron-positron colliders, and provided important input to the design of the detectors, the accelerator and the running program of such a machine. This would not have been possible without the intense interplay between theory and experiment and the long-term support by the SFB.

After the 12 years of this SFB, which saw the start-up of the LHC, the discovery of the Higgs boson, and, sadly, strong limits on further new particles, the physics progam of a Linear Collider is more important and timely than ever in the quest for widening the horizon of our knowledge.


\begin{footnotesize}

\bibliographystyle{sfb676}
\bibliography{sfb676_B1}

\end{footnotesize}


\end{document}

%% file: sfb676_macros_B1.tex
\usepackage{caption}
\usepackage{subcaption}
\usepackage[colorlinks=true, allcolors=blue]{hyperref}